\documentclass[letterpaper,12pt]{article}

\pdfoutput=1
\usepackage{graphicx}
\usepackage{amsmath}
\usepackage{hyperref}
\usepackage{bm}

\makeatletter
\renewcommand\section{\@startsection {section}{1}{\z@}%
{-3.5ex \@plus -1ex \@minus -0.2ex}%
{2.3ex \@plus 0.2ex}%
{\normalfont\normalsize\bfseries}}

\renewcommand\subsection{\@startsection{subsection}{2}{\z@}%
{-3.25ex \@plus -1ex \@minus -0.2ex}%
{1.5ex \@plus 0.2ex}%
{\normalfont\normalsize\bfseries}}

\def\@seccntformat#1{\csname the#1\endcsname.\quad}
\makeatother

\newcommand{\oline}[1]{\overline{\mkern-1.0mu#1\mkern0.0mu}}

\begin{document}

\setlength{\baselineskip}{3.75ex}

\noindent
\textbf{\LARGE Sharp hypotheses and organic fiducial}\\[2ex]
\textbf{\LARGE inference}\\[3ex]

\noindent
\textbf{Russell J. Bowater}\\
\emph{Independent researcher,
Doblado 110, Col.\ Centro, City of Oaxaca, C.P.\ 68000, Mexico.\\
Email address: as given on arXiv.org. Twitter profile:
\href{https://twitter.com/naked_statist}{@naked\_statist}\\ Personal website:
\href{https://sites.google.com/site/bowaterfospage}{sites.google.com/site/bowaterfospage}}
\\[2ex]

\noindent
\textbf{\small Abstract:}
{\small
A fundamental class of inferential problems are those characterised by there having been a
substantial degree of pre-data (or prior) belief that the value of a model parameter $\theta_j$ was
equal or lay close to a specified value\hspace{0.05em} $\theta^{*}_j$, which may, for example, be
the value that indicates the absence of a treatment effect or the lack of correlation between two
variables.
This paper puts forward a generally applicable `push-button' solution to problems of this type that
circumvents the severe difficulties that arise when attempting to apply standard methods of
inference, including the Bayesian method, to such problems.
Usually the only input of major note that is required from the user in implementing this solution
is the assignment of a pre-data or prior probability to the hypothesis that the parameter
$\theta_j$ lies in a narrow interval $[\theta_{j0},\theta_{j1}]$ that is assumed to contain the
value of interest $\theta^{*}_j$.
On the other hand, the end result that is achieved by applying this method is, conveniently, a
joint post-data distribution over all the parameters $\theta_1, \theta_2, \ldots, \theta_k$ of the
model concerned.
The proposed method is constructed by naturally combining a simple Bayesian argument with an
approach to inference called organic fiducial inference that was developed in a number of earlier
papers. To begin with, the main theoretical arguments underlying this combined Bayesian and
fiducial method are presented and discussed in detail. Various applications and useful extensions
of this methodology are then outlined in the latter part of the paper. The examples that are
considered are made relevant to the analysis of clinical trial data where appropriate.}
\\[3ex]
\textbf{\small Keywords:}
{\small Almost sharp hypothesis; Bayes factors; Bayesian analogies; Gibbs sampler; Incompatible
conditional distributions; Lack of pre-data knowledge; Model comparison; Organic fiducial
inference; Physical probability; Post-data predictive densities.}

\pagebreak
\section{Introduction}

The aim of this paper is to address a general type of problem in statistical inference that was
also the focus of attention in Bowater~(2019c). Therefore, let us begin in the same way that this
earlier paper began with a definition of a sharp and almost sharp hypothesis, and an introduction
to the specific problem of interest. For the moment, we will suppose that our overall aim is to
make inferences about an unknown parameter $\nu$ on the basis of a data set $x$ that was generated
by a sampling model that depends on the true value~of~$\nu$.

\vspace{3ex}
\noindent
\textbf{Definition 1: Sharp and almost sharp hypotheses}

\vspace{1ex}
\noindent
The hypothesis that the parameter $\nu$ lies in an interval $[\nu_0, \nu_1]$ will be defined as a
sharp hypothesis if $\nu_0 = \nu_1$, and as an almost sharp hypothesis if the difference $\nu_1 -
\nu_0$ is very small in the context of our general uncertainty about $\nu$ after the data $x$ have
been observed.

\vspace{3ex}
Clearly, any importance attached to a hypothesis of either of these two types should not generally
have a great effect on the way that we make inferences about $\nu$ on the basis of the data $x$ if
there had been no exceptional reason to believe that it would have been true or false before the
data were observed. Taking this into account, it will be assumed that we are in the following
scenario.

\vspace{3ex}
\noindent
\textbf{Definition 2: Scenario of interest}

\vspace{1ex}
\noindent
This scenario is characterised by there having been a substantial degree of belief before the data
were observed, i.e.\ a substantial pre-data belief, that a given sharp or almost sharp hypothesis
about the parameter $\nu$ could have been true, but if, on the other hand, this hypothesis had
been conditioned to be false, i.e.\ if $\nu$ had been conditioned not to lie in the interval
$[\nu_0, \nu_1]$, then there would have been very little or no pre-data knowledge about this
parameter over all of its allowable values outside of this interval.
In this scenario, the hypothesis in question will be referred to as the \emph{special} hypothesis.

\vspace{3ex}
Perhaps some may try to dismiss the importance of this type of scenario, however trying to make
data-based inferences about any given parameter of interest $\nu$ in such a scenario represents one
of the most fundamental problems of statistical inference that arise in practice. Let us consider
the following examples.

\vspace{3ex}
\noindent
\textbf{Example 1.1: Intervening in a system}

\vspace{1ex}
\noindent
If $\alpha$ is a parameter of one part of a system, and an intervention is made in a second part of
the system that is arguably completely disconnected from the first part, then there will be a high
degree of belief that the value of $\alpha$ will not change as a result of the intervention, i.e.\
there will be a strong belief in a sharp hypothesis about what will be the change in the value of
$\alpha$.

\vspace{3ex}
\noindent
\textbf{Example 1.2: A randomised-controlled trial}

\vspace{1ex}
\noindent
Let us imagine that a sample of patients is randomly divided into a group of $n_t$ patients, namely
the treatment group, that receive a new drug B, and a group of $n_c$ patients, namely the control
group, that receive a standard drug A.
We will assume that $e_t$ patients in the treatment group experience a given adverse event, e.g.\ a
heart attack, in a certain period of time following the start of treatment, and that $e_c$ patients
in the control group experience the same type of event in the same time period.
On the basis of this sample information, it will be supposed that the aim is to make inferences
about the relative risk $\pi_t/\pi_c$, where $\pi_t$ and $\pi_c$ are the population proportions of
patients who would experience the adverse event when given drug B and drug A respectively.
Now, if the action of drug B on the body is very similar to the action of drug A, which is in fact
often the case in practice when a new drug is being compared with a standard drug in this type of
clinical trial, then there may well have been a strong pre-data belief that this relative risk
would be close to one, or in other words, that the almost sharp hypothesis that the relative risk
would lie in a narrow interval containing the value one would be true.

\vspace{3ex}
It would appear that a common way to deal with there having been a strong pre-data belief that a
sharp or almost sharp hypothesis was true is to simply ignore the inconvenient presence of this
belief.
However, doing so means that inferences based on the observed data will often not be even remotely
honest. On the other hand, a formal method of addressing this issue that has received some
attention is the Bayesian method. We will now take a quick look at how this method would work in a
simple example.

\vspace{3ex}
\noindent
\textbf{Example 1.3: Application of the Bayesian method}

\vspace{1ex}
\noindent
Let us suppose that we are interested in making inferences about the mean $\mu$ of a normal
distribution that has a known variance $\sigma^2$ on the basis of a random sample $x$ of $n$ values
drawn from the distribution concerned. It will be assumed that we are in the scenario of
Definition~2 with the special hypothesis of interest being the sharp hypothesis that $\mu=0$.
Under the Bayesian paradigm, it would be natural to incorporate any degree of pre-data belief that
$\mu$ equals zero into the analysis of the data by assigning a positive prior probability to this
hypothesis.

However, the only generally accepted way of expressing a lack of pre-data knowledge about a model
parameter under this paradigm is the controversial strategy of placing a diffuse proper or improper
prior density over the parameter concerned. Taking this into account, let us assume, without a
great loss of generality, that the prior density function of $\mu$ conditional on $\mu \neq 0$ is a
normal density function with a mean of zero and a large variance $\sigma_{0}^2$.

The inadequacy of the strategy in question is clearly apparent in the uncertainty there would be in
choosing a value for the variance $\sigma_{0}^2$, and this issue becomes very hard to conceal after
appreciating that the amount of posterior probability given to\linebreak the hypothesis that
$\mu=0$ is highly sensitive to changes in this variance. For example, the natural desire to allow
the variance $\sigma_{0}^2$ to tend to infinity results in the posterior probability of this
hypothesis tending to one for any given data set $x$ and any given positive prior probability that
is assigned to $\mu$ equalling zero.

\vspace{3ex}
It can be easily argued, therefore, that the application of standard Bayesian theory to the case
just examined has an appalling outcome.
Moreover, applying the Bayesian strategy just described leads to outcomes of a similar type in
cases where the sampling density of the data given the parameter of interest $\nu$ is not normal,
and/or the prior density of this parameter has a more general form, and also, importantly, in cases
where the special hypothesis is an almost sharp rather than simply a sharp hypothesis.
This clearly gives us a strong motivation to look for an alternative method for making inferences
about $\nu$ in the scenario of interest.

One such method was proposed in Bowater and Guzm\'{a}n-Pantoja~(2019a) and further developed in
Bowater~(2019c), where it was referred to as being a special formalisation of what was called
bispatial inference.
In particular, this method is based on the interpretation of one-sided P values that correspond to
a composite null hypothesis that incorporates the special hypothesis of the scenario outlined in
Definition~2. No doubt many will find that this method has a certain degree of intuitive appeal.
However, the foundational justification given in Bowater~(2019c) as to why small P values should be
viewed as discrediting the type of null hypothesis in question was inadequate.
Also, it was not completely clear in Bowater~(2019c) how, by taking into account the size of the
type of one-sided P value under discussion, an appropriate post-data probability can in practice be
assigned to the null hypothesis to which it corresponds, which was one of the aims of the method
being referred to in this earlier paper.

In this paper, another type of method will be proposed for tackling the problem of inference in
question. It is reasonable to hope that this method will be regarded as representing a natural and
elegant way of addressing this problem of inference that is universally much better than the
approaches based on standard Bayesian inference and on the notion of bispatial inference that were
just mentioned.
In particular, it is a method that is based on using a simple Bayesian argument in combination with
a general approach to inference that was originally presented and named as organic fiducial
inference in Bowater~(2019b) and then further discussed in Bowater~(2020, 2021a) before being
clarified and modified in Bowater~(2021b).

Let us now briefly describe the structure of the paper. In the next five sections
(Sections~\ref{sec1} to~\ref{sec5}), the main theoretical arguments are presented for the method of
inference that will be advocated in the case where the sampling model depends on only one unknown
parameter. More specifically, the Bayesian component of the line of reasoning in question is
introduced and discussed in Section~\ref{sec4}, while a brief summary of organic fiducial inference
is given in Section~\ref{sec3}. These two approaches to inference are then brought together in
Sections~\ref{sec2} and~\ref{sec5} with the aim of solving the inferential problem of interest.
Examples of the application of the resulting method of inference are then presented in
Sections~\ref{sec6} and~\ref{sec8}.

Following on from this, a methodological adjustment is put forward in Section~\ref{sec7} that
enables undesirable discontinuities in the post-data density function of the parameter of interest
to be smoothed out. An example of the application of the adjusted method in question is then given
in Section~\ref{sec10}.

In the second part of the paper, two different ways of extending the originally proposed and the
adjusted method of inference just mentioned to the case where all model parameters are unknown are
developed.
More specifically, in Section~\ref{sec12}, a generalisation to address the case in question is
presented that is based on determining a joint post-data distribution over all the model parameters
by first constructing a set of full conditional post-data distributions for these parameters. An
example of the application of this extended methodology is then given in Section~\ref{sec14}. As an
alternative to this approach, a way of dealing with the case where all model parameters are unknown
is outlined in Section~\ref{sec15} that represents a more direct extension to the originally
proposed method for the case where only one parameter is unknown. Examples of the application of
this alternative methodology are then presented in Sections~\ref{sec16} and~\ref{sec18}.
The final section of the paper (Section~\ref{sec17}) contains some discussion of related approaches
to inference and an indication is given as to how the methodology that has been put forward could
be generalised further.

\vspace{3ex}
\section{Organic fiducial inference applied to the problem of interest}
\label{sec13}

\vspace{0.5ex}
\subsection{Overall assumptions}
\label{sec1}

Attention will be focused in the rest of this paper on a general version of the problem of
inference that was outlined in the Introduction. In particular, we will be interested in the
problem of making inferences about a set of parameters $\theta =\{\theta_i : i=1,2,\ldots,k\}$,
where each $\theta_i$ is a one-dimensional variable, on the basis of a data set
$x=\{x_i : i=1,2,\ldots,n\}$ that was generated by a given sampling model that is fully specified
by this set of parameters. Let the joint density or mass function of the data $x$ given the true
values of the parameters $\theta$ be denoted as $g(x\,|\,\theta)$.

For the moment, though, it will be assumed that the only unknown parameter on which the sampling
density $g(x\,|\,\theta)$ depends is the parameter $\theta_j$, either because there are no other
parameters in the model, or because all the other parameters are known.
Under this assumption, we will again suppose that the scenario of interest is the scenario outlined
in Definition~2.
To clarify, the special hypothesis in this scenario will be assumed to be the hypothesis that
the parameter $\theta_j$ lies in a given narrow interval, which we will now denote as the interval
$[\theta_{j0}, \theta_{j1}]$.

\vspace{3ex}
\subsection{A comparison of two Bayesian arguments}
\label{sec4}

As was the case in Bowater~(2020, 2021a, 2022), it will be assumed that the Bayesian approach to
inference is justified in any given situation by making an analogy between the uncertainty that
surrounds the validity of hypotheses that are relevant to the problem of inference being studied
and the uncertainty that surrounds the outcome of a well-understood physical experiment, e.g.\ the
outcome of randomly drawing a ball out of an urn of balls or the outcome of randomly spinning a
wheel. To clarify, specific outcomes of an experiment of this type and unions of these outcomes
will be regarded as having physical probabilities. What we exactly mean here by a physical
probability agrees with the definition of this type of probability that was given in
Bowater~(2022).

A full Bayesian analysis of the problem of inference outlined in Section~\ref{sec1} would require
that a prior density function $p(\theta_j)$ is placed over all values of the only unknown parameter
$\theta_j$.
Here an analogy would therefore need to be made between, on the one hand, our uncertainty, before
the data set $x$ is observed, both about the value of $\theta_j$ and the values that will make up
this data set, and on the other hand, our uncertainty about the outcome of a physical experiment
that is designed to generate a value of $\theta_j$ and the values in the data set $x$ by drawing
these values from the joint density of $\theta_j$ and $x$ that is specified by:
\vspace{0.5ex}
\[
p(\theta_j,x) = p(\theta_j)g(x\,|\,\theta_j)
\vspace{1ex}
\]
where $g(x\,|\,\theta_j)$ is the sampling density defined in Section~\ref{sec1}.

However, as discussed in the analysis of Example~1.3, it would arguably be very difficult to elicit
the prior density $p(\theta_j)$ under the assumption being made that we are in the scenario of
Definition~2.
Moreover, it can be argued that in this scenario, the prior density $p(\theta_j)$ conditional on
$\theta_j$ not lying in the interval $[\theta_{j0},\theta_{j1}]$ should be treated, in the case
where pre-data knowledge about $\theta_j$ outside of this interval is most lacking, as being a
completely unknown density function. Therefore, we may well be naturally led to consider dealing
with the problem at hand by performing a sensitivity analysis of the posterior density of the
parameter $\theta_j$ over all possible forms for the prior density $p(\theta_j)$ that are
consistent with a given prior probability being assigned to the hypothesis that $\theta_j$ lies in
the interval $[\theta_{j0},\theta_{j1}]$.

If this strategy was applied to Example~1.3 then, as was shown in Berger and Sellke~(1987), the
lower limit on the posterior probability of the hypothesis that $\theta_j$ lies in the interval
$[\theta_{j0},\theta_{j1}]$, i.e.\ the hypothesis that the mean $\mu$ equals zero in this example,
would be given by the expression:
\vspace{0.5ex}
\[
\left( 1 + \left( \frac{1-\mathtt{P}_0}{\mathtt{P}_0} \right)
\exp\left(\frac{\bar{x}^{2}}{2\sigma^2/n}\right) \right)^{-1}
\vspace{1ex}
\]
where $\mathtt{P}_0$ is the prior probability that $\mu$ equals zero and $\bar{x}$ is the sample
mean. This lower limit on the posterior probability in question is achieved when the prior density
of $\mu$, apart from placing a probability mass of $\mathtt{P}_0$ at the point $\mu=0$, also places
a probability mass of $(1 - \mathtt{P}_0)$ at the point $\mu = \bar{x}$.
On the other hand, as was the case for the strategy outlined in Example~1.3, the upper limit on the
posterior probability of the mean $\mu$ being equal to zero would be equal to one, and therefore it
could be argued that the current strategy being considered could not, in general, be regarded as
being practically viable.

Nevertheless, in searching for an alternative strategy we do not need to move completely away from
the Bayesian paradigm. In particular, let us make an analogy of the type just discussed in which
the prior density of $\theta_j$ conditioned on $\theta_j$ not lying in the interval $[\theta_{j0},
\theta_{j1}]$ is essentially treated as though it does not exist. However, the precise analogy that
we need to make depends on the scenario in which we find ourselves.
Therefore, let us give a more specific definition of the scenario of interest by assuming that not
only are we in the scenario of Definition~2, but that if, before the data $x$ were observed, the
parameter $\theta_j$ had been conditioned to lie in the interval $[\theta_{j0},\theta_{j1}]$, then
very little or nothing would have been known about where in this interval the parameter $\theta_j$
would lie.

Under these assumptions, we will make an analogy between, on the one hand, our pre-data uncertainty
both about the value of $\theta_j$ and the values that will make up the data set $x$, and on the
other hand, our uncertainty about the outcome of a physical experiment that will randomly select
either a value $\theta_{jA}$ or a value $\theta_{jB}$ \pagebreak for the parameter
$\theta_j$\hspace{0.05em}, and then will generate the data $x$ by substituting this chosen value of
$\theta_j$ into the formula of the sampling density $g(x\,|\,\theta_j)$.
The key detail that establishes this analogy as being a generally acceptable one to make is that it
will be assumed that we know that the value $\theta_{jA}$ lies in the interval
$[\theta_{j0},\theta_{j1}]$ and that the value $\theta_{jB}$ lies outside of this interval, but
apart from this, the values $\theta_{jA}$ and $\theta_{jB}$ will be assumed to be completely
unknown to us. To clarify, in conducting the physical experiment of interest, the joint density of
$\theta_j$ and the values in the data set $x$ would be given by:
\begin{equation}
\label{equ1}
p(\theta_j,x) = p(\theta_j)g(x\,|\,\theta_j)
\end{equation}
in which
\vspace{1.5ex}
\begin{equation}
\label{equ8}
p(\theta_j) = \left\{
\begin{array}{ll}
p_0(\theta_{jA}) & \mbox{if $\theta_j = \theta_{jA}$}\\[1ex]
p_0(\theta_{jB}) = 1 - p_0(\theta_{jA})\, & \mbox{if $\theta_j = \theta_{jB}$}\\[1ex]
0 & \mbox{otherwise}
\end{array}
\right.
\vspace{2.5ex}
\end{equation}
where $p_0(\theta_{jA})$ is a specified probability.
Given our lack of information concerning the values of $\theta_{jA}$ and $\theta_{jB}$, the
probability $p_0(\theta_{jA})$ is in effect our prior probability that $\theta_j$ lies in the
interval $[\theta_{j0},\theta_{j1}]$, while $p_0(\theta_{jB})$ is in effect our prior probability
that $\theta_j$ lies outside of this interval.

However, in using the analogy that has just been outlined, we can not form the posterior
probability of $\theta_j$ lying in the interval $[\theta_{j0},\theta_{j1}]$ and the posterior
probability of $\theta_j$ not lying in this interval by simply conditioning the joint density of
$\theta_j$ and the data $x$ given in equation~(\ref{equ1}) on the data $x$ that is actually
observed. This is because, as a consequence of the values $\theta_{jA}$ and $\theta_{jB}$ being
unknown, the sampling densities $g(x\,|\,\theta_{jA})$ and $g(x\,|\,\theta_{jB})$ will also be
unknown. Nevertheless, it is perfectly reasonable to contemplate estimating the densities
$g(x\,|\,\theta_{jA})$ and $g(x\,|\,\theta_{jB})$ using all the information that is available to
us. As will be detailed in Section~\ref{sec2}, we will choose to estimate these densities using
organic fiducial inference. Before that though, with the aim of giving an adequate background to
this later section, we will present, in the next section, a brief general overview of the theory of
organic fiducial inference.

\pagebreak
\subsection{An overview of organic fiducial inference}
\label{sec3}

As mentioned in the Introduction, organic fiducial inference is a theory of inference that was
developed in Bowater~(2019b, 2020, 2021b). Also, precursory work that forms part of the basis of
this theory can be found in Bowater~(2017, 2018). For a complete account of this theory of
inference, it is recommended that Bowater~(2019b) is studied with careful attention given to the
modifications to this theory that are presented in Bowater~(2021b).
In this section, we will simply give a brief overview of the methodology that underlies organic
fiducial inference in the case where the parameter $\theta_j$ is the only unknown parameter in the
sampling model, which is of course the case of current interest.

\vspace{3ex}
\noindent
\textbf{Fiducial statistics}

\vspace{1ex}
\noindent
Let $V(x)$ be a set of univariate statistics that together form a sufficient or approximately
sufficient set of statistics for the parameter $\theta_j$. To clarify, a set of statistics will be
defined as being an approximately sufficient set of statistics for $\theta_j$ if conditioning the
distribution of the data set of interest $x$ on this set of statistics leads to a distribution of
this data set $x$ that does not depend heavily on the value of the parameter $\theta_j$.

Now, if a given set $V(x)$, as just defined, only contains one statistic that is not an ancillary
statistic, then that statistic will be called the \emph{fiducial statistic} $Q(x)$ of this set
$V(x)$. Assuming that the set $V(x)$ is of this nature, we will denote all the statistics in $V(x)$
except for the statistic $Q(x)$ as the set of statistics\hspace{0.05em} $U(x)\hspace{-0.05em}
=\hspace{-0.05em} \{U_i(x): i=1,2,\ldots,m\}$, and we will refer to the statistics in this set as
being the ancillary complements of $Q(x)$.
Note that often it may be possible to find a set $V(x)$ that consists of just a single
suf\-fi\-cient statistic for $\theta_j$, and of course in these cases, the fiducial statistic
$Q(x)$ will be this sufficient statistic and the set $U(x)$ will be empty.

\vspace{3ex}
\noindent
\textbf{Data generating algorithm}

\vspace{1ex}
\noindent
Independent of the way in which the data set $x$ was actually generated, it will be assumed that
this data set was generated by a specific algorithm called the \emph{data generating algorithm}.

After generating the values $u$ of the ancillary complements $U(x)$, if any exist, of a given
fiducial statistic $Q(x)$, this algorithm proceeds by determining a value $q(x)$ for this latter
statistic by setting it equal to a function $\varphi$ that depends on the parameter $\theta_j$, the
values $u$ and an already generated value $\gamma$ of a variable $\Gamma$ that is referred to as
the \emph{primary random variable} (primary r.v.). The density function of $\Gamma$, which we will
denote as the density $\pi_0(\gamma)$, does not depend on the parameter $\theta_j$. In the final
step of the algorithm the data set $x$ is randomly drawn from the sampling density or mass function
$g(x\,|\,\theta_1,\theta_2, \ldots, \theta_k)$ conditioned on the statistics $Q(x)$ and $U(x)$
being equal to their already generated values.

\vspace{3ex}
\noindent
\textbf{Types of fiducial argument}

\vspace{1ex}
\noindent
In the theory of organic fiducial inference, the fiducial argument, which is usually considered to
be a single argument, is broken down into three sub-arguments referred to as the strong, moderate
and weak fiducial arguments.

\vspace{3ex}
\noindent
\textbf{Strong or standard fiducial argument}

\vspace{1ex}
\noindent
This is the argument that the density function of the primary r.v.\ $\Gamma$ after the data have
been observed, i.e.\ the post-data density function of $\Gamma$, should be equal to the pre-data
density function of $\Gamma$, i.e.\ the density $\pi_0(\gamma)$.

\vspace{3ex}
\noindent
\textbf{Moderate fiducial argument}

\vspace{1ex}
\noindent
It will be assumed that this argument is only applicable if, on observing the data $x$, there
exists some positive measure set of values of the primary r.v.\ $\Gamma$ over which the pre-data
density function $\pi_0(\gamma)$ was positive, but over which the post-data density function of
$\Gamma$, which will be denoted as the density function $\pi_1(\gamma)$, is necessarily zero.
Under this condition, it is the argument that, over the set of values of $\Gamma$ for which the
density function $\pi_1(\gamma)$ is necessarily positive, the relative height of this function
should be equal to the relative height of the density function $\pi_0(\gamma)$, i.e.\ the heights
of these two functions should be proportional.

\vspace{3ex}
\noindent
\textbf{Weak fiducial argument}

\vspace{1ex}
\noindent
This argument will be assumed to be only applicable if neither the strong nor the moderate fiducial
argument is considered to be acceptable. By using this argument, the most appropriate post-data
density of $\Gamma$, i.e.\ the density $\pi_1(\gamma)$, can be determined in the most general type
of scenario, i.e.\ determined with regard to the most general form of a function of $\theta_j$
specified before the data were observed that is referred to as the \emph{global pre-data function}
of $\theta_j$. Let us now define this latter function.

\vspace{3ex}
\noindent
\textbf{Global pre-data (GPD) function}

\vspace{1ex}
\noindent
The global pre-data (GPD) function $\omega_G(\theta_j)$ is used to express pre-data knowledge, or a
lack of such knowledge, about the unknown parameter $\theta_j$.
More specifically, this function may be any given non-negative and upper bounded function of the
parameter $\theta_j$. It is a function that only needs to be specified up to a proportionality
constant, in the sense that, if it is multiplied by a positive constant, then the value of the
constant is redundant.

\vspace{3ex}
\noindent
\textbf{Local pre-data (LPD) function}

\vspace{1ex}
\noindent
The local pre-data (LPD) function $\omega_L(\theta_j)$ is also used to express pre-data knowledge
about the parameter $\theta_j$, but in a way that is different from the way this knowledge is
expressed by the GPD function. More specifically, it may be any given non-negative function of the
parameter $\theta_j$ that is locally integrable over the space of this parameter.
Similar to a GPD function, a LPD function only needs to be specified up to a proportionality
constant.

The role of the LPD function is to facilitate the completion of the definition of the joint
post-data density function of the primary r.v.\ $\Gamma$ and the parameter $\theta_j$ in cases
where using either the strong or moderate fiducial argument alone is not sufficient to achieve
this. For this reason, the LPD function is in fact redundant in many situations.

\vspace{3ex}
\noindent
\textbf{Principles for defining univariate fiducial density functions}

\vspace{1ex}
\noindent
Given the data $x$, there are two mutually consistent principles in the theory being considered for
defining the fiducial density function of the parameter $\theta_j$ conditional on all other
parameters $\theta_{-j}=\{\theta_1,\ldots,\theta_{j-1},\theta_{j+1},\ldots,\theta_k\}$ being known,
i.e.\ the density function $f(\theta_j\,|\,\theta_{-j},x)$. Let us briefly discuss these two
principles.

\pagebreak
\noindent
\textbf{Principle 1 for defining the density $f(\theta_j\,|\,\theta_{-j},x)$}

\vspace{1ex}
\noindent
Loosely speaking, this principle applies if, after the data have been observed, the function
$\varphi$ that forms part of the data generating algorithm outlined earlier defines a bijective
mapping between the set of all possible values of the primary r.v.\ $\Gamma$, which we will denote
as the set $G_x$, and the set of all possible values of the parameter $\theta_j$.
Under this condition, the fiducial density $f(\theta_j\,|\,\theta_{-j},x)$ is directly defined by
the post-data density of $\Gamma$, i.e.\ the density $\pi_1(\gamma)$, which is assumed to be given
by the following expression:
\vspace{1ex}
\begin{equation}
\label{equ2}
\pi_1(\gamma) = \left\{
\begin{array}{ll}
\mathtt{C}_0\hspace{0.1em} \omega_G(\theta_j(\gamma))\hspace{0.05em} \pi_0(\gamma)\, & \mbox{if
$\gamma \in G_x$}\\[1ex]
0 & \mbox{otherwise}
\end{array}
\right.
\vspace{1ex}
\end{equation}
in which $\theta_j(\gamma)$ is the value of the parameter $\theta_j$ that maps onto the value
$\gamma$ of the variable $\Gamma$ according to the definition of the function $\varphi$, and
where\hspace{0.05em} $\omega_G(\theta_j(\gamma))$ is the GPD function of $\theta_j$ defined earlier
and $\mathtt{C}_0$ is a normalising constant.

\vspace{3ex}
\noindent
\textbf{Principle 2 for defining the density $f(\theta_j\,|\,\theta_{-j},x)$}

\vspace{1ex}
\noindent
This principle can often be applied in situations where, after the data have been observed, various
values of the parameter $\theta_j$ are consistent, according to the definition of the function
$\varphi$, with any given value of the primary r.v.\ $\Gamma$.
The main step that underlies this principle is to form a joint density of $\Gamma$ and $\theta_j$
that is based on the marginal density of $\Gamma$ being the post-data density $\pi_1(\gamma)$
defined by equation~(\ref{equ2}) and the conditional density of $\theta_j$ given that $\Gamma$
equals a specific value $\gamma$ being determined using the Bayesian paradigm on the basis of the
pre-data information that is expressed by what was earlier defined to be a LPD function
$\omega_L(\theta_j)$. The result of marginalising this joint density of $\Gamma$ and $\theta_j$
with respect to the variable $\Gamma$ is then quite naturally regarded as being the fiducial
density $f(\theta_j\,|\,\theta_{-j},x)$.

\vspace{3ex}
\noindent
\textbf{An extension to Principle 2}

\vspace{1ex}
\noindent
In situations where neither Principle~1 nor the basic version of Principle~2 just referred to can
be applied, the use of an extended version of Principle~2 was advocated in Sections~7.2 and~8 of
Bowater~(2019b). To calculate the fiducial density $f(\theta_j\,|\,\theta_{-j},x)$ using this
extension to Principle~2, we first apply the version of Principle~2 \pagebreak just summarised to
determine a preliminary version of this fiducial density of $\theta_j$ that would be appropriate in
a general scenario where nothing or very little was known about the parameter $\theta_j$ before the
data were observed. Let this preliminary version of the fiducial density in question be denoted as
$f_{S}(\theta_j\,|\,\theta_{-j},x)$. The fiducial density of $\theta_j$ that is actually of concern
to us, i.e.\ the density $f(\theta_j\,|\,\theta_{-j},x)$, is then given by the expression:
\vspace{0.5ex}
\begin{equation}
\label{equ7}
f(\theta_j\,|\,\theta_{-j},x) = \mathtt{C}_1\hspace{0.05em}
\omega_G(\theta_j)f_{S}(\theta_j\,|\,\theta_{-j},x)
\vspace{0.5ex}
\end{equation}
where $\omega_G(\theta_j)$ is the GPD function of $\theta_j$ that corresponds to the genuine
scenario of interest, and $\mathtt{C}_1$ is a normalising constant. Observe that if, in this
definition of the fiducial density $f(\theta_j\,|\,\theta_{-j},x)$, the preliminary fiducial
density $f_{S}(\theta_j\,|\,\theta_{-j},x)$ was derived using Principle~1 rather than Principle~2,
then the resulting density $f(\theta_j\,|\,\theta_{-j},x)$ ought to be equivalent to what it would
have been if it had been derived directly by using Principle~1.

\vspace{3ex}
\subsection{Estimating the densities \texorpdfstring{$g(x\,|\,\theta_{jA})$}{g(x | theta jA)} and
\texorpdfstring{$g(x\,|\,\theta_{jB})$}{g(x | theta jB)}}
\label{sec2}

Let us return to the problem of interest that was last discussed in Section~\ref{sec4}.
In particular, we now will consider the question of how to estimate the densities
$g(x\,|\,\theta_{jA})$ and $g(x\,|\,\theta_{jB})$. First, let us recall that these two sampling
densities would be known if the values of $\theta_{jA}$ and $\theta_{jB}$ were known. Second, since
in the Bayesian analogy being used the values of $\theta_{jA}$ and $\theta_{jB}$ are, by contrast,
assumed to be known, the door is not closed on the possibility of using the data $x$ to estimate
the densities $g(x\,|\,\theta_{jA})$ and $g(x\,|\,\theta_{jB})$, which to clarify, means estimating
the sampling density $g(x\,|\,\theta_{j})$ under the assumption that $\theta_j$ lies in the
interval $[\theta_{j0},\theta_{j1}]$ and that $\theta_j$ does not lie in this interval,
respectively.
Therefore, let us explore the idea of estimating the two sampling densities $g(x\,|\,\theta_{jA})$
and $g(x\,|\,\theta_{jB})$ using the post-data predictive density of a future data set $x'$ that is
the same size as the observed data set $x$ under the assumption that $\theta_j=\theta_{jA}$ and
that $\theta_j=\theta_{jB}$, respectively.

We should point out, though, that it would be unwise to try to derive the post-data predictive
density of the data $x'$ under each of these two assumptions by using the Bayesian paradigm.
The first reason for this is that, as has already \pagebreak been discussed, it is very difficult
to use prior densities of the parameter $\theta_j$ to express the type of pre-data knowledge that
we would have about $\theta_j$ in the scenario of interest under the two assumptions in question,
i.e.\ to express that nothing or very little would have been known about $\theta_j$ before the data
were observed if $\theta_j$ had either been conditioned to lie inside or outside of the interval
$[\theta_{j0},\theta_{j1}]$.
The second reason for not using the Bayesian paradigm to perform the current task is that the
analogies that would need to be made to justify the use of this paradigm in constructing the
post-data predictive densities in question would effectively imply that the overall Bayesian
analogy that would be used would be the same as the first type of Bayesian analogy detailed in
Section~\ref{sec4}. In other words, the way being presently considered of tackling the problem of
inference outlined in Section~\ref{sec1} would be effectively reduced to the type of full Bayesian
analysis of this problem that was discussed in Section~\ref{sec4}.
In an analysis of this type, the overall prior density of $\theta_j$ that is implied by the
underlying assumptions that are made would clearly need to be updated to an overall posterior
density of $\theta_j$ by means of a standard application of Bayes' theorem.
The major drawbacks of conducting such a full Bayesian analysis of the problem of interest have, of
course, already been carefully laid out (see the Introduction and Section~\ref{sec4}).

To construct an appropriate post-data predictive density of the data $x'$ either under the
condition that $\theta_j=\theta_{jA}$ or that $\theta_j=\theta_{jB}$, we will therefore consider
using the theory of organic fiducial inference that was summarised in the previous section.
This approach has the advantage that it offers a natural way in which a lack of pre-data knowledge
about the parameter $\theta_j$ can be expressed when $\theta_j$ is either conditioned to lie inside
or outside of the interval $[\theta_{j0},\theta_{j1}]$, and also, the use of this approach does not
conflict with the Bayesian analogy that we are making as part of the overall strategy that is being
adopted, i.e.\ the second type of Bayesian analogy detailed in Section~\ref{sec4}.

The two post-data predictive densities of the data $x'$ that we need to determine, which are now
more specifically the fiducial predictive densities of the data $x'$ that result from separately
applying the condition that $\theta_j=\theta_{jA}$ and that $\theta_j=\theta_{jB}$, will be denoted
as being the density functions $f(x'\,|\,\theta_{jA},\theta_{-j},x)$ and
$f(x'\,|\,\theta_{jB},\theta_{-j},x)$, respectively. In broad terms, these two fiducial predictive
densities are defined as follows:
\vspace{1ex}
\begin{equation}
\label{equ3}
f(x'\,|\,\theta_{jA},\theta_{-j},x) =
\int^{\hspace{0.03em} \mbox{\footnotesize $\theta_{j1}$}}_{\hspace{0.03em}
\mbox{\footnotesize $\theta_{j0}$}} g(x'\,|\,\theta) f(\theta_{j}\,|\,\theta_{jA},\theta_{-j},x)
d\theta_{j}
\end{equation}
and
\begin{eqnarray}
f(x'\,|\,\theta_{jB},\theta_{-j},x) & = &
\int^{\hspace{0.03em}\mbox{\footnotesize $\theta_{j0}$}}_{-\infty} g(x'\,|\,\theta)
f(\theta_{j}\,|\,\theta_{jB},\theta_{-j},x) d\theta_{j}
\nonumber \\[1ex]
\label{equ4}
& & + \hspace{0.2em}
\int^{\infty}_{\hspace{0.03em}\mbox{\footnotesize $\theta_{j1}$}} g(x'\,|\,\theta)
f(\theta_{j}\,|\,\theta_{jB},\theta_{-j},x) d\theta_{j}
\end{eqnarray}
\par \vspace{1ex} \noindent
where $g(x'\,|\,\theta)$ is the general sampling density as defined in Section~\ref{sec1} and the
densities $f(\theta_{j}\,|\,\theta_{jA},\theta_{-j},x)$ and
$f(\theta_{j}\,|\,\theta_{jB},\theta_{-j},x)$ are the fiducial densities of $\theta_j$ under the
condition that $\theta_j$ lies in the interval $[\theta_{j0},\theta_{j1}]$ and that $\theta_j$ does
not lie in this interval, respectively.

Since the sampling density $g(x'\,|\,\theta)$ is known, we only need to determine the fiducial
densities $f(\theta_{j}\,|\,\theta_{jA},\theta_{-j},x)$ and
$f(\theta_{j}\,|\,\theta_{jB},\theta_{-j},x)$ in the expressions just given to be able to obtain
the predictive densities of the data $x'$ that are of interest.
In doing this, let us first observe that as $\theta_j=\theta_{jA}$ implies that $\theta_j$ must lie
in the interval $[\theta_{j0}, \theta_{j1}]$, it follows that, with regard to determining the
fiducial density $f(\theta_{j}\,|\,\theta_{jA},\theta_{-j},x)$, the GDP function
$\omega_G(\theta_j)$ must be equal to zero outside of the interval in question.
On the other hand, as $\theta_j=\theta_{jB}$ implies that $\theta_j$ must lie outside of the
interval $[\theta_{j0}, \theta_{j1}]$, it follows that, with regard to determining the fiducial
density $f(\theta_{j}\,|\,\theta_{jB},\theta_{-j},x)$, the GDP function $\omega_G(\theta_j)$ must
be equal to zero inside the interval in question.
Given that it has been assumed that nothing or very little would have been known about the
parameter $\theta_j$ before the data $x$ were observed if $\theta_j$ had either been conditioned to
lie inside or outside of the interval $[\theta_{j0}, \theta_{j1}]$, it is quite natural to specify
the two GPD functions for $\theta_j$ that are of interest so that they are equal to a positive
constant over the intervals where we know that they are not equal to zero.
Therefore, we will define the GPD function of $\theta_j$ as:
\vspace{0.5ex}
\begin{equation}
\label{equ5}
\omega_G(\theta_j) = \left\{
\begin{array}{ll}
a\ & \mbox{if $\theta_j \in [\theta_{j0}, \theta_{j1}]$}\\[1ex]
0 & \mbox{otherwise}
\end{array}
\right.
\end{equation}
and as:
\vspace{2ex}
\begin{equation}
\label{equ6}
\omega_G (\theta_j) = \left\{
\begin{array}{ll}
a\ & \mbox{if $\theta_j \in (-\infty,\theta_{j0}) \cup (\theta_{j1},\infty)$}\\[1ex]
0 & \mbox{otherwise}
\end{array}
\right.
\pagebreak
\end{equation}
with regard to the construction of the fiducial densities
$f(\theta_{j}\,|\,\theta_{jA},\theta_{-j},x)$ and $f(\theta_{j}\,|\,\theta_{jB},$ $\theta_{-j},x)$,
respectively, where $a$ is any given positive constant.
Note that the two GPD functions just defined would be regarded as being \emph{neutral} GPD
functions according to the terminology of Bowater~(2019b).

If Principle~1 summarised in the previous section can be applied, then using this principle on the
basis of the GPD functions of $\theta_j$ given in equations~(\ref{equ5}) and~(\ref{equ6}) would
imply that the fiducial densities $f(\theta_{j}\,|\,\theta_{jA},\theta_{-j},x)$ and
$f(\theta_{j}\,|\,\theta_{jB},\theta_{-j},x)$ would, in general, be derived by using the moderate
fiducial argument as defined earlier. Furthermore, if Principle~1, Principle~2 or the extension to
Principle~2 described in Section~\ref{sec3} can be applied, then these fiducial densities would be
defined by the general expression for the fiducial density $f(\theta_j\,|\,\theta_{-j},x)$ given in
equation~(\ref{equ7}).

Observe that, if the fiducial densities $f(\theta_{j}\,|\,\theta_{jA}, \theta_{-j},x)$ and
$f(\theta_{j}\,|\,\theta_{jB}, \theta_{-j},x)$ are determined by Principle~2 or the extension to
Principle~2 in question, then a LPD function for $\theta_j$ would be required to determine the
fiducial density $f_{S}(\theta_{j}\,|\,\theta_{-j},x)$ in equation~(\ref{equ7}).
However, since this latter fiducial density needs to be derived under the assumption that there was
no or very little pre-data knowledge about the parameter $\theta_j$, it may be difficult to specify
a single LPD function for $\theta_j$ that is representative of our pre-data beliefs about
$\theta_j$.
For this reason, it may often be of benefit to analyse how making different but nevertheless
sensible choices for the LPD function of $\theta_j$ affects the form of the density
$f_{S}(\theta_{j}\,|\,\theta_{-j},x)$, and therefore in turn, the forms of the densities
$f(\theta_{j}\,|\,\theta_{jA},\theta_{-j},x)$ and $f(\theta_{j}\,|\,\theta_{jB},\theta_{-j},x)$.

\vspace{3ex}
\subsection{Returning to the Bayesian analogy of interest}
\label{sec5}

Under the assumptions that define the physical experiment on which the Bayesian analogy of current
interest is based, i.e.\ the second type of Bayesian analogy detailed in Section~\ref{sec4}, the
posterior distribution of the parameter $\theta_j$ is obtained by conditioning the joint
distribution of $\theta_j$ and the data $x$ that is defined by equations~(\ref{equ1})
and~(\ref{equ8}) on the data $x$ that is actually observed. In particular, doing this leads to the
posterior probabilities that $\theta_j=\theta_{jA}$ and that $\theta_j=\theta_{jB}$ \pagebreak
being defined as:
\begin{equation}
\label{equ9}
p(\theta_j = \theta_{jA}\,|\,\theta_{-j},x) =
\frac{p_0(\theta_{jA})g(x\,|\,\theta_{jA})}{p_0(\theta_{jA})g(x\,|\,\theta_{jA}) +
p_0(\theta_{jB})g(x\,|\,\theta_{jB})}
\end{equation}
and
\[
p(\theta_j = \theta_{jB}\,|\,\theta_{-j},x) = 1 - p(\theta_j = \theta_{jA}\,|\,\theta_{-j},x)
\vspace{1ex}
\]
respectively, where the notation being used here is the same as used in equations~(\ref{equ1})
and~(\ref{equ8}).

On substituting the fixed but unknown sampling densities $g(x\,|\,\theta_{jA})$ and
$g(x\,|\,\theta_{jB})$ in equation~(\ref{equ9}) by the estimates of these densities that were put
forward in the previous section, i.e.\ the fiducial predictive density of the future data set $x'$
in the cases where $\theta_j=\theta_{jA}$ and where $\theta_j=\theta_{jB}$, the posterior
probabilities of the hypotheses that $\theta_j=\theta_{jA}$ and that $\theta_j=\theta_{jB}$ just
given, which were derived purely under the Bayesian paradigm, become simply justifiable post-data
probabilities of these hypotheses that are \vspace{1ex} defined as:
\begin{equation}
\label{equ10}
\tilde{p}(\theta_j = \theta_{jA}\,|\,\theta_{-j},x) =
\frac{p_0(\theta_{jA}) f(x' = x\,|\,\theta_{jA},\theta_{-j},x)}
{p_0(\theta_{jA})f(x' = x\,|\,\theta_{jA},\theta_{-j},x) +
p_0(\theta_{jB})f(x' = x\,|\,\theta_{jB},\theta_{-j},x)}
\vspace{1ex}
\end{equation}
and
\begin{equation}
\label{equ22}
\tilde{p}(\theta_j = \theta_{jB}\,|\,\theta_{-j},x) =
1 - \tilde{p}(\theta_j = \theta_{jA}\,|\,\theta_{-j},x)
\vspace{1.5ex}
\end{equation}
respectively, where the densities $f(x'\,|\,\theta_{jA},\theta_{-j},x)$ and
$f(x'\,|\,\theta_{jB},\theta_{-j},x)$ are as defined in equations~(\ref{equ3}) and~(\ref{equ4}).
Observe that, according to the definitions of the values $\theta_{jA}$ and $\theta_{jB}$ that we
have been using, the probability\hspace{0.05em} $\tilde{p}(\theta_j\hspace{-0.05em} =
\theta_{jA}\,|\,\theta_{-j},x)$ is, in effect, our post-data probability that $\theta_j$ lies in
the interval $[\theta_{j0},\theta_{j1}]$ and the probability $\tilde{p}(\theta_j =
\theta_{jB}\,|\,\theta_{-j},x)$ is, in effect, our post-data probability that $\theta_j$ does not
lie in this interval.

To obtain a post-data density function of $\theta_j$ over the whole of the real line it is
perfectly consistent with the line of reasoning being currently adopted that the post-data
densities of $\theta_j$ that result from separately applying the condition that $\theta_j$ lies in
the interval $[\theta_{j0},\theta_{j1}]$ and that $\theta_j$ does not lie in this interval are,
respectively, the fiducial densities $f(\theta_{j}\,|\,\theta_{jA},\theta_{-j},x)$ and
$f(\theta_{j}\,|\,\theta_{jB},\theta_{-j},x)$ defined in the previous section.
Given this natural assumption, it is clear that the complete definition of the post-data density of
$\theta_j$ is given by the expression:
\[
\tilde{p}(\theta_j\,|\,\theta_{-j},x) = f(\theta_{j}\,|\,\theta_{jA},\theta_{-j},x)\hspace{0.05em}
\tilde{p}(\theta_j = \theta_{jA}\,|\,\theta_{-j},x) +
f(\theta_{j}\,|\,\theta_{jB},\theta_{-j},x)\hspace{0.05em}
\tilde{p}(\theta_j = \theta_{jB}\,|\,\theta_{-j},x)
\]
or equivalently by the expression:
\vspace{0.5ex}
\begin{eqnarray}
\tilde{p}(\theta_j\,|\,\theta_{-j},x) & = & f(\theta_{j}\,|\,\theta_j \in
[\theta_{j0}, \theta_{j1}], \theta_{-j},x)\hspace{0.05em} \tilde{p}(\theta_j \in
[\theta_{j0}, \theta_{j1}]\,|\,\theta_{-j},x) \nonumber \\
\label{equ11}
& & + \hspace{0.2em}
f(\theta_j\,|\,\theta_j \notin [\theta_{j0}, \theta_{j1}],\theta_{-j},x)\hspace{0.05em}
\tilde{p}(\theta_j \notin [\theta_{j0}, \theta_{j1}]\,|\,\theta_{-j},x)
\end{eqnarray}

\vspace{3ex}
\section{First example: Inference about a normal mean with variance known}
\label{sec6}

To give a first example of the application of the method of inference outlined in the previous
sections, let us apply this method to the problem of inference referred to in Example~1.3 in the
Introduction, i.e.\ the problem of making inferences about a normal mean $\mu$ when the population
variance $\sigma^2$ is known.

With the aim of providing a context for this problem, let us imagine that a random sample of $n$
patients are being constantly monitored with regard to the concentration of a certain chemical in
their blood.
In this scenario, the data value $x_i$ will be assumed to be simply the measurement of this
concentration for the $i$th patient at a time point $\mathtt{t}_2$ minus the same type of
measurement for this patient taken at a time point $\mathtt{t}_1$, where the time point
$\mathtt{t}_1$ is immediately before the patient is subjected to some kind of intervention and the
time point $\mathtt{t}_2$ is immediately after this intervention. Furthermore, we will suppose that
the data values $x_i$ are normally distributed around a population mean $\mu$, and also that the
sample size $n$ is large enough such that it is acceptable to use the sample vari-\linebreak ance
$s^2$ as a substitute for the population variance $\sigma^2$.

Finally, it will be imagined that the intervention in question is not expected to affect by very
much or at all the concentration of the chemical of interest for each of the patients concerned.
Therefore, a substantial degree of pre-data belief would exist that the population mean change
$\mu$ in this concentration in going from the time point $\mathtt{t}_1$ to the time point
$\mathtt{t}_2$ will be very small.
More specifically, it will be assumed that we find ourselves in the scenario of Definition~2 with
the parameter of interest in this scenario now being $\mu$ and with the special hypothesis being
that $\mu$ lies in the interval $[-\varepsilon, \varepsilon]$, where $\varepsilon$ is a small
non-negative constant.

As clearly the sample mean $\bar{x}$ is a sufficient statistic for $\mu$, it can therefore be
assumed to be the fiducial statistic $Q(x)$ in applying the theory of organic fiducial inference as
summarised in Section~\ref{sec3}. Based on this assumption, the function $\varphi$ that forms part
of the data generating algorithm described in Section~\ref{sec3} can be expressed as:
\[
\varphi(\Gamma,\mu)=\mu+(\sigma/\sqrt{n}\hspace{0.1em})\hspace{0.05em}\Gamma
\]
where the primary r.v.\ $\Gamma \sim \mbox{N}(0,1)$, which means that it is being assumed that the
fiducial statistic $\bar{x}$ is determined by setting $\Gamma$ equal to its already generated value
$\gamma$ in the formula: $\bar{x} = \varphi(\Gamma,\mu)$.

If, with regard to the construction of the fiducial densities $f(\theta_{j}\,|\,\theta_{jA},
\theta_{-j},x)$ and\linebreak $f(\theta_{j}\,|\,\theta_{jB},\theta_{-j}, x)$, the GPD functions of
$\mu$ are defined according to equations~(\ref{equ5}) and~(\ref{equ6}) respectively, then these
fiducial densities will be derived in the present case under Principle~1 described in
Section~\ref{sec3} by using the moderate fiducial argument.
As a result, it turns out that the fiducial densities in question, i.e.\ the fiducial densities
$f(\mu\,|\,\mu \in [-\varepsilon, \varepsilon],x)$ and
$f(\mu\,|\,\mu \notin [-\varepsilon, \varepsilon],x)$ in the present case, are obtained by
conditioning the normal density of $\mu$ given by:
\[
\mu \sim \mbox{N}(\bar{x}, \sigma^2/n)
\vspace{0.5ex}
\]
on the event that $\mu \in [-\varepsilon, \varepsilon]$ and the event that $\mu \notin
[-\varepsilon, \varepsilon]$, respectively.

Taking into account that the sampling density of every value $x_i$ in the data set $x$, i.e.\ the
density $g(x_i\,|\,\theta)$ for $i=1,2,\ldots,n$, is a normal density with mean $\mu$ and variance
$\sigma^2$, the fiducial predictive densities $f(x'\,|\,\theta_{jA},\theta_{-j},x)$ and
$f(x'\,|\,\theta_{jB},\theta_{-j},x)$, i.e.\ the densities
$f(x'\,|\,\mu \in [-\varepsilon,\varepsilon],x)$ and
$f(x'\,|\,\mu \notin [-\varepsilon,\varepsilon],x)$ in the present case, are defined by the
expressions in equations~(\ref{equ3}) and~(\ref{equ4}).
On eliciting a value for the prior probability $p_0(\theta_{jA})$, which in the present case is the
prior probability that $\mu\hspace{-0.05em} \in [-\varepsilon,\varepsilon]$, we can, as a next
step, use the expression in equation~(\ref{equ10}) to determine the post-data probability
$\tilde{p}(\theta_j = \theta_{jA}\,|\,\theta_{-j},x)$, i.e.\ the post-data probability that
$\mu\hspace{-0.05em} \in [-\varepsilon,\varepsilon]$.
Just to clarify, under the assumptions of the current example, this post-data probability is
defined as follows:\\[2ex]
$\tilde{p}(\mu \in [-\varepsilon, \varepsilon]\,|\,x) =$
\begin{flalign}
\label{equ12}
&& \frac{p(\mu \in [-\varepsilon,\varepsilon]) f(x'= x \,|\,\mu \in [-\varepsilon,\varepsilon],x) }
{p(\mu \in [-\varepsilon,\varepsilon]) f(x'= x\,|\,\mu \in [-\varepsilon,\varepsilon],x)  +
(1 - p(\mu \in [-\varepsilon,\varepsilon])) f(x'= x\,|\,\mu \notin [-\varepsilon,\varepsilon],x)}
\raisetag{-1.5ex}
\end{flalign}
where $p(\mu \in [-\varepsilon,\varepsilon])$ is the prior probability that $\mu \in
[-\varepsilon,\varepsilon]$.

Observe that the fiducial predictive densities $f(x'\,|\,\mu \in [-\varepsilon,\varepsilon],x)$
and $f(x'\,|\,\mu \notin [-\varepsilon,\varepsilon],x)$ will, in general, need to be computed on
the basis of a numerical approximation. However, let us take a look at the simple case where
$\varepsilon=0$, which means that the fiducial predictive densities in question are now of course
the densities $f(x'\,|\,\mu = 0,x)$ and $f(x'\,|\,\mu \neq 0,x)$. In relation to this case, it can
easily be established that the fiducial predictive densities of a future sample mean
$\bar{x}\hspace{0.05em}'$ that result from applying either the condition that $\mu=0$ or that $\mu
\neq 0$, i.e.\ the densities $f(\bar{x}\hspace{0.05em}'\,|\,\mu = 0,x)$ and
$f(\bar{x}\hspace{0.05em}'\,|\,\mu \neq 0,x)$, are given by the analytical expressions:
\[
\bar{x}\hspace{0.05em}'\,|\,\mu = 0,x \sim \mbox{N} (0, \sigma^2/n)
\vspace{-1ex}
\]
and
\[
\bar{x}\hspace{0.05em}'\,|\,\mu \neq 0,x \sim \mbox{N} (\bar{x}, 2\sigma^2/n)
\vspace{0.5ex}
\]
respectively. On taking into account that the values that make up the future data set $x'$ are
independent normal random values, it can be appreciated that these expressions effectively specify
the predictive densities $f(x'\,|\,\mu = 0,x)$ and $f(x'\,|\,\mu \neq 0,x)$.
We should point out that, since the predictive density $f(x'\,|\,\mu = 0,x)$ is derived under the
precise condition that $\mu = 0$, it is simply the sampling density $g(x\,|\,\mu = 0)$ with the
data $x'$ taking the place of the data $x$.

As a next step, let us substitute the predictive densities $f(x'\,|\,\mu = 0,x)$ and
$f(x'\,|\,\mu \neq 0,x)$ just defined into equation~(\ref{equ12}). As a result, we obtain the
following analytical expression for the post-data probability that $\mu$ equals zero:
\vspace{1ex}
\begin{equation}
\label{equ13}
\tilde{p}(\mu = 0 \,|\,x) = \frac{\raisebox{0.3ex}{$p(\mu = 0)$}}
{\raisebox{-0.3ex}{$p(\mu = 0) + (1 - p(\mu = 0))(1/\sqrt{2})\exp((n/2)(\bar{x}/\sigma)^2)$}}
\vspace{1ex}
\end{equation}
where $p(\mu = 0)$ is the prior probability that $\mu$ equals zero.

To illustrate the general use of the calculation that appears in equation~(\ref{equ12}), Figure~1
shows various plots of the post-data probability $\tilde{p}(\mu \in [-\varepsilon,
\varepsilon]\,|\,x)$ against the observed mean $\bar{x}$.
This figure is based on the assumption that the variance $\sigma^2$ is equal to the sample size
$n$, i.e.\ it is being assumed that the standard error\hspace{0.05em}
$\sigma/\sqrt{n}$\hspace{0.05em} of the sample mean $\bar{x}$ is equal to one, and for the sake of
convenience, this figure is only concerned with the case where $\bar{x} \geq 0$, which will be
assumed to also be the case of interest \pagebreak in the following discussion.
The short-dashed, long-dashed and solid curves that form the three upper curves in Figure~1
correspond to the scenario where the prior probability $p(\mu \in [-\varepsilon,\varepsilon])$ is
equal to 0.5, while the three lower curves in this figure correspond to the scenario where the
prior probability $p(\mu \in [-\varepsilon,\varepsilon])$ is equal to 0.3.
The two solid curves in Figure~1 represent plots of $\tilde{p}(\mu \in [-\varepsilon,
\varepsilon]\,|\,x)$ against $\bar{x}$ in the case where $\varepsilon = 0$, and are therefore, in
effect, plots of the function given in equation~(\ref{equ13}) under the assumptions that are
currently being made.
On the other hand, the two long-dashed curves in this figure correspond to the case where
$\varepsilon = 0.25$, while the two short-dashed curves correspond to the case where
$\varepsilon = 0.5$.

\begin{figure}[t]
\begin{center}
\includegraphics[width=5in]{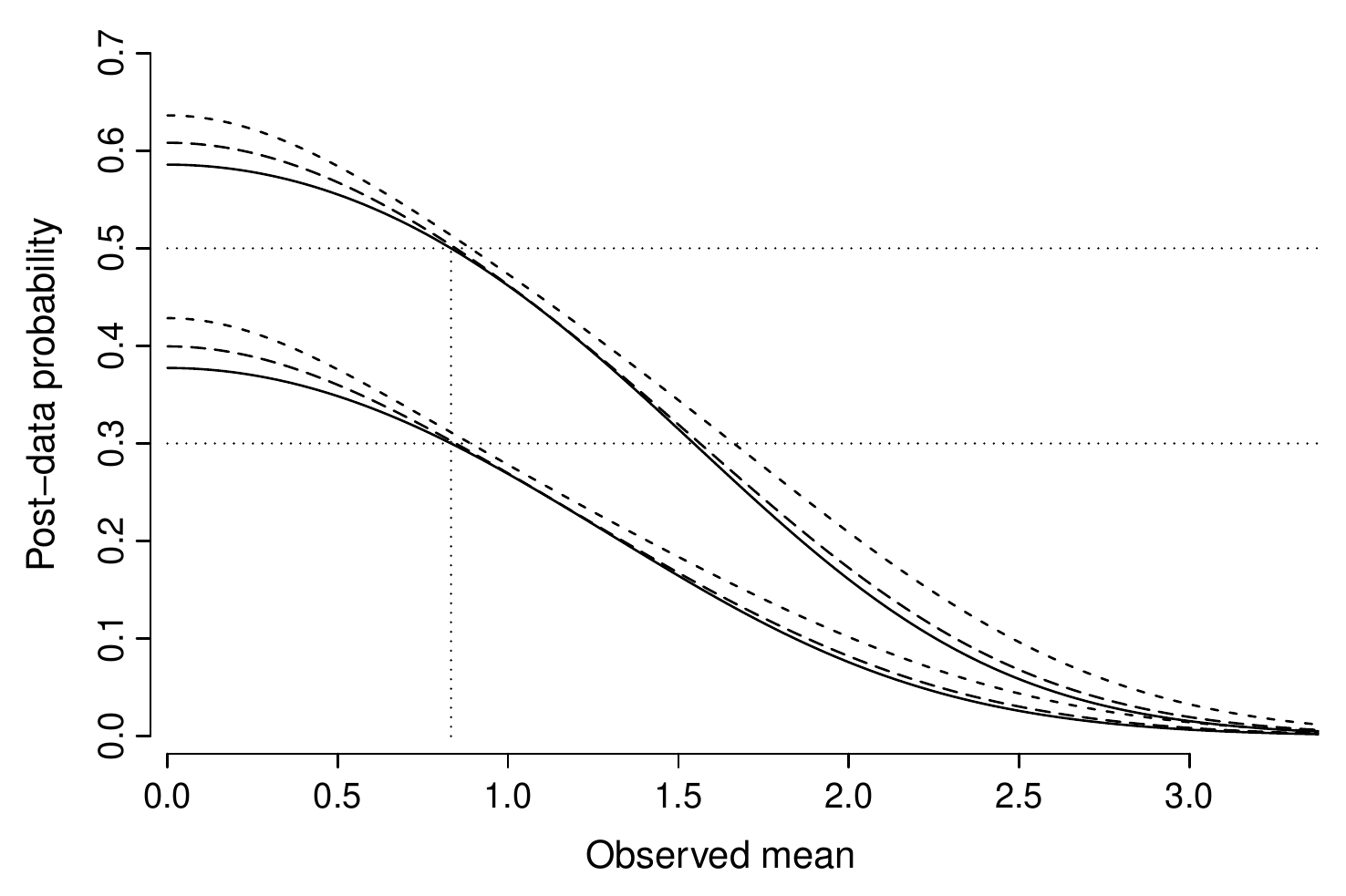}
\caption{\small{Post-data probability of a normal mean $\mu$ lying in the interval $[-\varepsilon,
\varepsilon]$ against the sample mean $\bar{x}$ for three values of $\varepsilon$ (namely 0, 0.25
and 0.5) and two values for the prior probability of $\mu \in [-\varepsilon, \varepsilon]$ (namely
0.3 and 0.5)}}
\end{center}
\end{figure}

Given that setting $\varepsilon$ equal to 0.25 implies that the interval $[-\varepsilon,
\varepsilon]$ has half the width of the standard error of the sample mean, a realistic choice of
$\varepsilon$ would arguably be less than 0.25 since, according to the scenario of Definition~2,
the width of the interval of values for $\mu$ that are permitted under the special hypothesis about
$\mu$ should be very small in the context of our post-data uncertainty about $\mu$.
However, if we had set $\varepsilon$ to be equal to say 0.1, the plots of the post-data probability
$\tilde{p}(\mu \in [-\varepsilon, \varepsilon]\,|\,x)$ against the observed mean $\bar{x}$ would
have been almost indistinguishable from the solid curves in Figure~1 for the two values of the
prior probability $p(\mu \in [-\varepsilon,\varepsilon])$ being considered.
The reason why plots of $\tilde{p}(\mu \in [-\varepsilon, \varepsilon]\,|\,x)$ against $\bar{x}$
for the case where $\varepsilon = 0.5$ have been included in Figure~1, i.e.\ the plots traced out
by the short-dashed curves in this figure, is to show what happens when $\varepsilon$ is chosen to
be equal to a value that arguably is unrealistically large.

It can be gathered from Figure~1 that not only do large values of the sample mean $\bar{x}$ lead
to the post-data probability of the hypothesis that $\mu \in [-\varepsilon, \varepsilon]$ being
less than the prior probability of this hypothesis, but small values of the sample mean $\bar{x}$
lead to the reverse being true. For example, in the case where $\varepsilon=0$, the post-data
probability of the hypothesis that $\mu \in [-\varepsilon, \varepsilon]$ is greater than the prior
probability of this hypothesis for values of $\bar{x}$ that are less than 0.8325, independent of
the choice that is made for the prior probability in question.
Also, in the case where $\varepsilon=0$, it is of interest to note that for the three values of the
sample mean $\bar{x}$ that correspond to the two-sided P value of the Z test of the null hypothesis
that $\mu = 0$ being 0.05, 0.01 and 0.001, the post-data probability $\tilde{p}(\mu = 0 \,|\,x)$
would be 0.1716, 0.0488 and 0.0063, respectively, if the prior probability that $\mu = 0$ was 0.5,
and would be 0.0816, 0.0215 and 0.0027, respectively, if this prior probability was 0.3.

Under the assumptions that have already been made, we can not only define, for any given
$\varepsilon$ and any given prior probability $p(\mu \in [-\varepsilon,\varepsilon])$, the
post-data probability $\tilde{p}(\mu \in [-\varepsilon, \varepsilon]\,|\,x)$, but we can also
define the post-data density function of $\mu$ over the whole of the real line according to the
expression in equation~(\ref{equ11}).
Just to clarify, under the assumptions of the present example, this post-data density function is
defined as follows:
\vspace{-0.5ex}
\begin{equation}
\label{equ14}
\tilde{p}(\mu\,|\,x) =  f(\mu\,|\,\mu \in [-\varepsilon, \varepsilon],x)\hspace{0.05em}
\tilde{p}(\mu \in [-\varepsilon, \varepsilon]\,|\,x) +
f(\mu\,|\,\mu \notin [-\varepsilon, \varepsilon]\,|\,x)\hspace{0.05em} \tilde{p}(\mu \notin
[-\varepsilon, \varepsilon]\,|\,x)
\vspace{1ex}
\end{equation}

To illustrate the use of the calculation in this equation, Figure~2 shows plots of the post-data
density $\tilde{p}(\mu\,|\,x)$ for various values of the observed mean $\bar{x}$.
All plots in this fig\-ure correspond to the case where $\varepsilon=0.1$ and where the prior
probability $p(\mu \in [-0.1,$ $0.1])$ is equal to 0.3, and again it has been assumed that the
variance $\sigma^2$ is equal to the sample size $n$.
The short-dashed, long-dashed and solid curves in Figure~2 correspond to the observed mean
$\bar{x}$ being equal to 1.7, 2.1 and 2.5, respectively.
In keeping with the results that were reported in Figure~1, it can be seen from the present figure
that for the range of values of the sample mean $\bar{x}$ being considered, the post-data
probability $\tilde{p}(\mu \in [-0.1, 0.1]\,|\,x)$ decreases fairly sharply as $\bar{x}$
increases.

\begin{figure}[t]
\begin{center}
\includegraphics[width=5.5in]{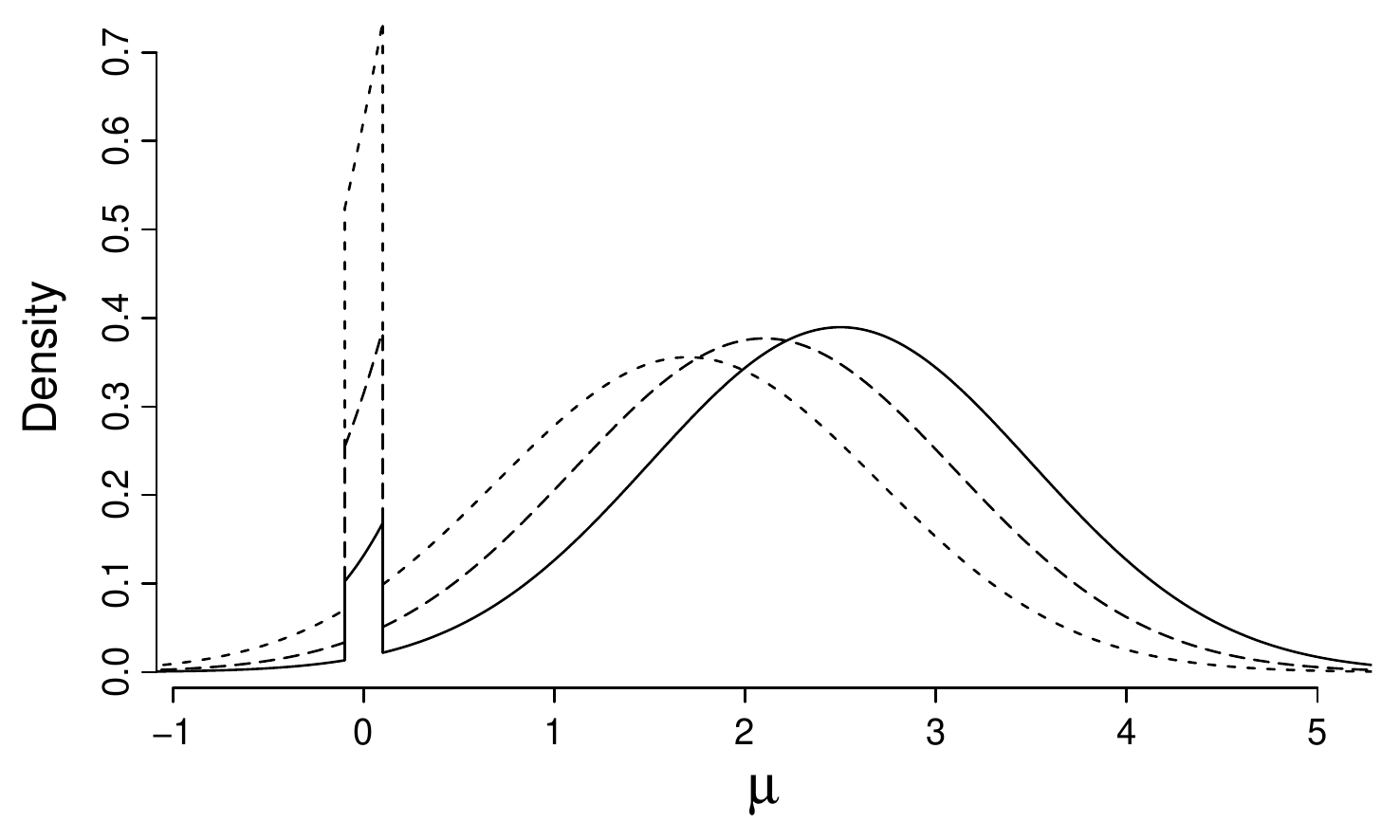}
\caption{\small{Post-data densities of a normal mean $\mu$ for three different values of the sample
mean $\bar{x}$ (namely 1.7, 2.1 and 2.5) in the case where $\varepsilon=0.1$ and the prior
probability that $\mu \in [-0.1,0.1]$ is equal to 0.3}}
\end{center}
\end{figure}

Let us now take a closer look at how sensitive the post-data density $\tilde{p}(\mu\,|\,x)$ is to
changes in the value of $\varepsilon$ when the prior probability $p(\mu \in [-\varepsilon,
\varepsilon])$ is held constant. In this regard, Figure~3 shows plots of the post-data probability
of $\mu$ lying in the fixed interval $[-0.2, 0.2]$, i.e.\ the probability
$\tilde{p}(\mu \in [-0.2, 0.2]\,|\,x)$, against the observed mean $\bar{x}$ for three different
values of $\varepsilon$. More specifically, the solid, short-dashed and long-dashed curves in this
figure correspond to the cases where $\varepsilon$ is equal to 0, 0.1 and 0.2, re\-spec\-tive\-ly.
Similar to Figure~1, the three upper curves in Figure~3 correspond to the case where the prior
probability $p(\mu \in [-\varepsilon,\varepsilon])$ is equal to 0.5, while the three lower curves
in this figure correspond to the case where the prior probability $p(\mu \in [-\varepsilon,
\varepsilon])$ is equal to 0.3. Again it has been assumed that $\sigma^2 = n$.

\begin{figure}[t]
\begin{center}
\includegraphics[width=4.75in]{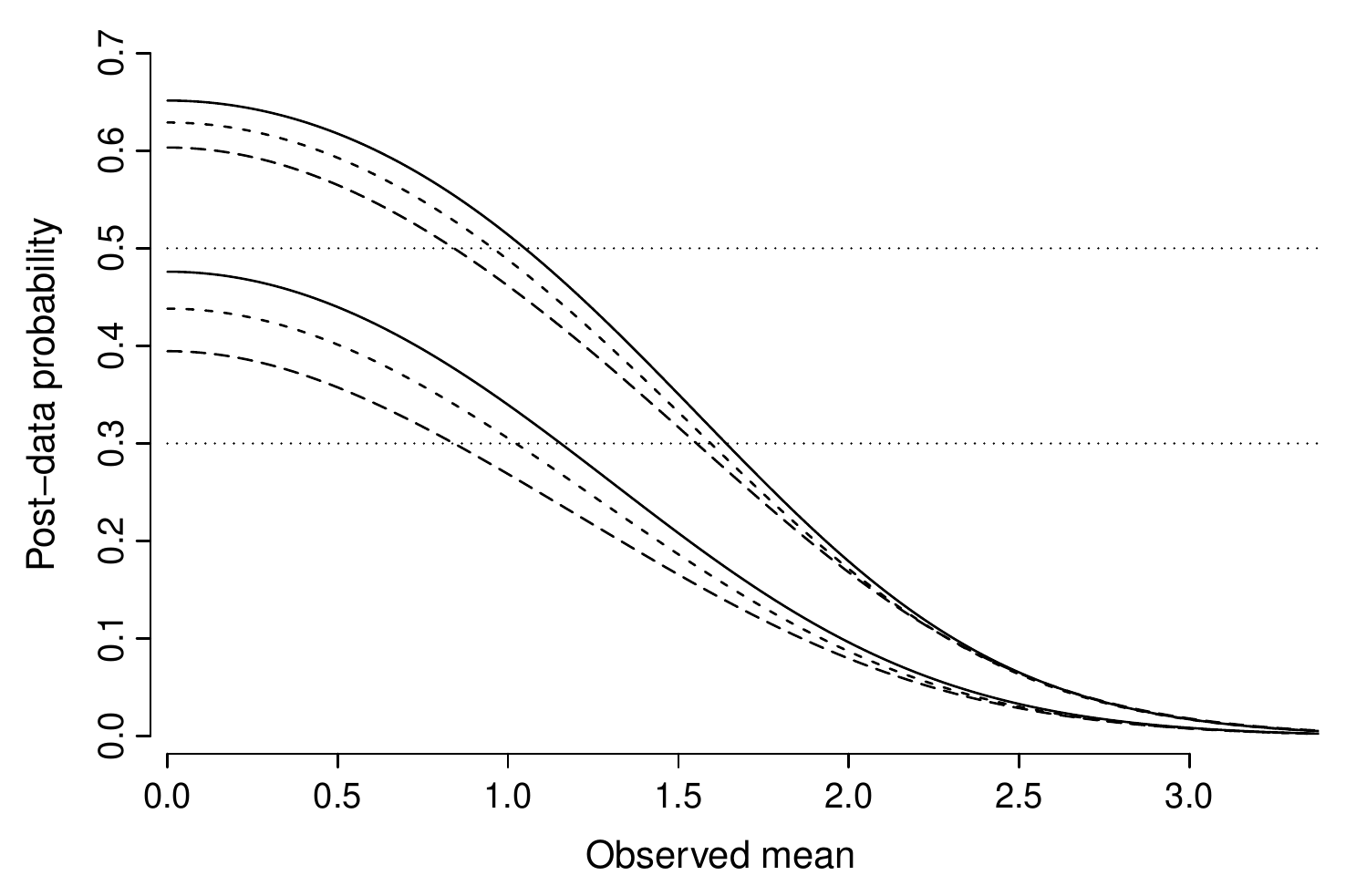}
\caption{\small{Post-data probability of a normal mean $\mu$ lying in the fixed interval $[-0.2,
0.2]$ against the sample mean $\bar{x}$ for three values of $\varepsilon$ (namely 0, 0.1 and 0.2)
and two values for the prior probability of $\mu \in [-\varepsilon, \varepsilon]$ (namely 0.3 and
0.5)}}
\end{center}
\end{figure}

Taking into account the results reported in Figure~3, we can conclude that if we are interested in
determining a post-data probability for $\mu$ being, in some sense, `close' to zero on the basis of
the post-data density $\tilde{p}(\mu\,|\,x)$ defined in equation~(\ref{equ14}), then there will be
some degree of insensitivity to the choice made for the value of $\varepsilon$.
However, in trying to achieve this particular goal, it will be generally the case that a reasonable
amount of care will nevertheless need to be shown in choosing an appropriate value for
$\varepsilon$. In this regard, the precise choice made for $\varepsilon$ will become a more
important issue as the prior probability $p(\mu \in [-\varepsilon, \varepsilon])$ becomes smaller.

\vspace{3ex}
\section{Second example: Inference about a binomial proportion}
\label{sec8}

Let us imagine that a random sample of patients are switched from being given a standard drug A to
being given a new drug B. After a period of time has passed, they are asked which out of the two
drugs A and B they prefer. Let $x$ denote the number of patients who prefer drug B to drug A and
let $n$ denote the number of patients left in the sample after patients who do not express a
preference have been excluded. Given the sample proportion $x/n$, we will suppose that the aim is
to make inferences about what this proportion would be in the population, which will be denoted as
the proportion $\pi$, under the standard assumption that, before the experiment took place, the
probability of observing any given count $x$ was specified by the binomial mass function in this
case, i.e.\ the function:
\vspace{0.5ex}
\begin{equation}
\label{equ15}
g_0(x\,|\,\pi) = \binom{n}{x} \pi^{\hspace{0.05em}x} (1-\pi)^{n-x}\ \ \ \mbox{for $x=0,1,\ldots,n$}
\vspace{1.5ex}
\end{equation}
For a similar reason with regard to the nature of drugs A and B as that given in Example~1.2 of the
Introduction, let us also assume that the scenario of Definition~2 applies with the parameter of
interest in this scenario now being the proportion $\pi$ and with the special hypothesis being that
$\pi$ lies in a narrow interval centred at 0.5, which will be denoted as $[0.5-\varepsilon,
0.5+\varepsilon]$.

As clearly the observed count $x$ is a sufficient statistic for the proportion $\pi$, it can
therefore be assumed to be the fiducial statistic $Q(x)$ in this example.
If it is also supposed that the primary r.v.\ $\Gamma$ has a uniform distribution over the interval
$(0,1)$, then the function $\varphi$ that forms part of the data generating algorithm described
in Section~\ref{sec3} can be expressed as:
\begin{equation}
\label{equ38}
\varphi(\Gamma,\pi)= \min \left\{ y: \Gamma < \mbox{\Large $\sum$}_{\mbox{\footnotesize
$j\hspace{-0.25em}=\hspace{-0.25em}0$}}^{\hspace{0.05em}\mbox{\footnotesize $y$}}\hspace{0.3em}
g_0(j\,|\,\pi) \right\}
\vspace{1ex}
\end{equation}
where the mass function $g_0(j\,|\,\pi)$ is as defined in equation~(\ref{equ15}), which means that
it is being assumed that the sample count $x$ is determined by setting $\Gamma$ equal to its
already generated value $\gamma$ in the formula $x = \varphi(\Gamma,\pi)$.

On defining the GPD function of $\pi$ according to equations~(\ref{equ5}) and~(\ref{equ6}) in the
cases where $\theta_j=\theta_{jA}$ and $\theta_j=\theta_{jB}$ respectively, the fiducial densities
$f(\theta_{j}\,|\,\theta_{jA},\theta_{-j},x)$ and $f(\theta_{j}\,|\,\theta_{jB},\theta_{-j},x)$
referred to in Sections~\ref{sec2} and~\ref{sec5} are then derived in the present case using the
extension to Principle~2 that was outlined in Section~\ref{sec3}.
As explained in Section~\ref{sec2}, to apply this extension to Principle~2, a LPD function of $\pi$
is required so that the fiducial density $f_{S}(\theta_{j}\,|\,\theta_{-j},x)$ in
equation~(\ref{equ7}) can be determined. To give an example, let us assume that this LPD function
is defined as follows:
\vspace{1ex}
\[
\omega_{L}(\pi) = \left\{
\begin{array}{ll}
b\ & \mbox{if $0 \leq \pi \leq 1$}\\[1ex]
0 & \mbox{otherwise}
\end{array}
\right.
\vspace{0.5ex}
\]
where $b$ is a positive constant.

By applying the extension to Principle~2 being referred to, we now can define the fiducial
densities $f(\theta_{j}\,|\,\theta_{jA},\theta_{-j},x)$ and $f(\theta_{j}\,|\,\theta_{jB},
\theta_{-j},x)$, which in the present case are the densities $f(\pi\,|\,\pi \in [0.5-\varepsilon,
0.5+\varepsilon], x)$ and $f(\pi\,|\,\pi \notin [0.5-\varepsilon, 0.5+\varepsilon], x)$, in the
following way:
\vspace{0.5ex}
\[
f(\pi\,|\,\pi \in [0.5-\varepsilon, 0.5+\varepsilon], x) = \left\{
\begin{array}{ll}
\mathtt{C}_2 \mbox{\footnotesize $\displaystyle \int$}_{\hspace{-0.3em}0}^1 \omega_*(\pi \,|\,
\gamma) d\gamma\, & \mbox{if $\pi \in [0.5-\varepsilon, 0.5+\varepsilon]$}\\[1.75ex]
0 & \mbox{otherwise}
\end{array}
\right.
\vspace{1ex}
\]
\[
f(\pi\,|\,\pi \notin [0.5-\varepsilon, 0.5+\varepsilon], x) = \left\{
\begin{array}{ll}
\mathtt{C}_3 \mbox{\footnotesize $\displaystyle \int$}_{\hspace{-0.3em}0}^1 \omega_*(\pi \,|\,
\gamma) d\gamma\, & \mbox{if $\pi \notin [0.5-\varepsilon, 0.5+\varepsilon]$}\\[1.75ex]
0 & \mbox{otherwise}
\end{array}
\right.
\vspace{2.5ex}
\]
where the conditional density function $\omega_*(\pi \,|\, \gamma)$ is given by:
\vspace{1.5ex}
\[
\omega_*(\pi\,|\,\gamma) = \left\{
\begin{array}{ll}
\mathtt{C}_4\hspace{-0.03em}(\gamma)\hspace{0.1em}\pi^{\hspace{0.05em}x} (1-\pi)^{n-x}\, &
\mbox{if $\pi \in \pi(\gamma)$}\\[1ex]
0 & \mbox{otherwise}
\end{array}
\right.
\vspace{1ex}
\]
in which $\pi(\gamma)$ is the set of values of $\pi$ that map onto the value $\gamma$ for the
primary r.v.\ $\Gamma$ according to the equation $x = \varphi(\Gamma,\pi)$ given the observed value
of $x$, and where $\mathtt{C}_2$, $\mathtt{C}_3$ and $\mathtt{C}_4(\gamma)$ are normalising
constants, the latter of which clearly must depend on the value of~$\gamma$.

Taking into account that the sampling mass function of the count $x$ is as defined in
equation~(\ref{equ15}), the fiducial predictive densities $f(x'\,|\,\theta_{jA},\theta_{-j},x)$ and
$f(x'\,|\,\theta_{jB},\theta_{-j},x)$, i.e.\ the densities $f(x'\,|\,\pi \in [0.5-\varepsilon,
0.5+\varepsilon], x)$ and $f(x'\,|\,\pi \notin [0.5-\varepsilon, 0.5+\varepsilon], x)$ in the
present case, are given by the expressions in equations~(\ref{equ3}) and~(\ref{equ4}).
On eliciting a value for the prior probability $p_0(\theta_{jA})$, which in the present case is the
prior probability that $\pi \in [0.5-\varepsilon,0.5+\varepsilon]$, we can, as a next step, use the
expression in equation~(\ref{equ10}) to determine the post-data probability $\tilde{p}(\theta_j =
\theta_{jA}\,|\,\theta_{-j},x)$, i.e.\ the post-data probability that $\pi \in [0.5-\varepsilon,
0.5+\varepsilon]$.
Just to clarify, under the assumptions of the current example, this post-data probability is
defined \pagebreak as follows:\\
$\tilde{p}(\pi \in [0.5 - \varepsilon, 0.5 + \varepsilon]\,|\,x) =$
\begin{flalign*}
&& \frac{\mathtt{P}_0\hspace{0.05em}f(x'= x \,|\,\pi \in [0.5 -\varepsilon,0.5 + \varepsilon],x)}
{\mathtt{P}_0\hspace{0.05em}f( x'= x\,|\,\pi \in [0.5-\varepsilon, 0.5+\varepsilon],x)  +
(1 - \mathtt{P}_0) f( x'= x\,|\,\pi \notin [0.5-\varepsilon, 0.5+\varepsilon],x)}
\end{flalign*}
\par \vspace{1ex} \noindent
where $\mathtt{P}_0$ is the prior probability that $\pi \in [0.5-\varepsilon, 0.5+\varepsilon]$.
Finally, using this result, the post-data density function of $\pi$ over all allowable values for
$\pi$, i.e.\ over the interval $[0,1]$, can be determined according to the expression in
equation~(\ref{equ11}), which means that, under the assumptions of the present example, this
density function is defined as follows:
\begin{eqnarray*}
\tilde{p}(\pi\,|\,x) & = & f(\pi \,|\, \pi \in
[0.5 - \varepsilon, 0.5 + \varepsilon], x)\hspace{0.05em} \tilde{p}(\pi \in [0.5 - \varepsilon,
0.5 + \varepsilon] \,|\, x)\\ & & +
\hspace{0.2em}
f(\pi\,|\,\pi \notin [0.5 - \varepsilon, 0.5 + \varepsilon], x)\hspace{0.05em}
\tilde{p}(\pi \notin [0.5 - \varepsilon, 0.5 + \varepsilon] \,|\, x)
\end{eqnarray*}

To illustrate the use of the calculations just outlined, Figures~4(a) to~4(c) show some results
from generating values from the post-data density $\tilde{p}(\pi\,|\,x)$ defined by the equation
just given in three different scenarios. In particular, the histograms in these figures represent
the density $\tilde{p}(\pi\,|\,x)$ in these three scenarios.
The general assumptions that were made to perform the calculations that underlie the figures in
question were that $\varepsilon=0.01$, the sample size $n$ is equal to 16 and the prior probability
of $\pi$ lying in the interval $[0.49,0.51]$ is equal to 0.3. More specifically, the histogram in
Figure~4(a) represents results that were obtained in the case where the observed count $x$ was set
equal to 5, while the histograms in Figures~4(b) and~4(c) represent results that were obtained when
$x$ was set equal to 4 and 3, respectively.
The numerical output on which these histograms are based was generated by the method of importance
sampling, more specifically by appropriately weighting, in each case, a sample of two million
independent random values drawn from the fiducial density $f_{S}(\theta_{j}\,|\,\theta_{-j},x)$ or,
more precisely, the density $f_{S}(\pi\,|\,x)$ referred to in the above discussion that
corresponded to the particular case concerned.

\begin{figure}[t]
\begin{center}
\makebox[\textwidth]{\includegraphics[width=7.5in]{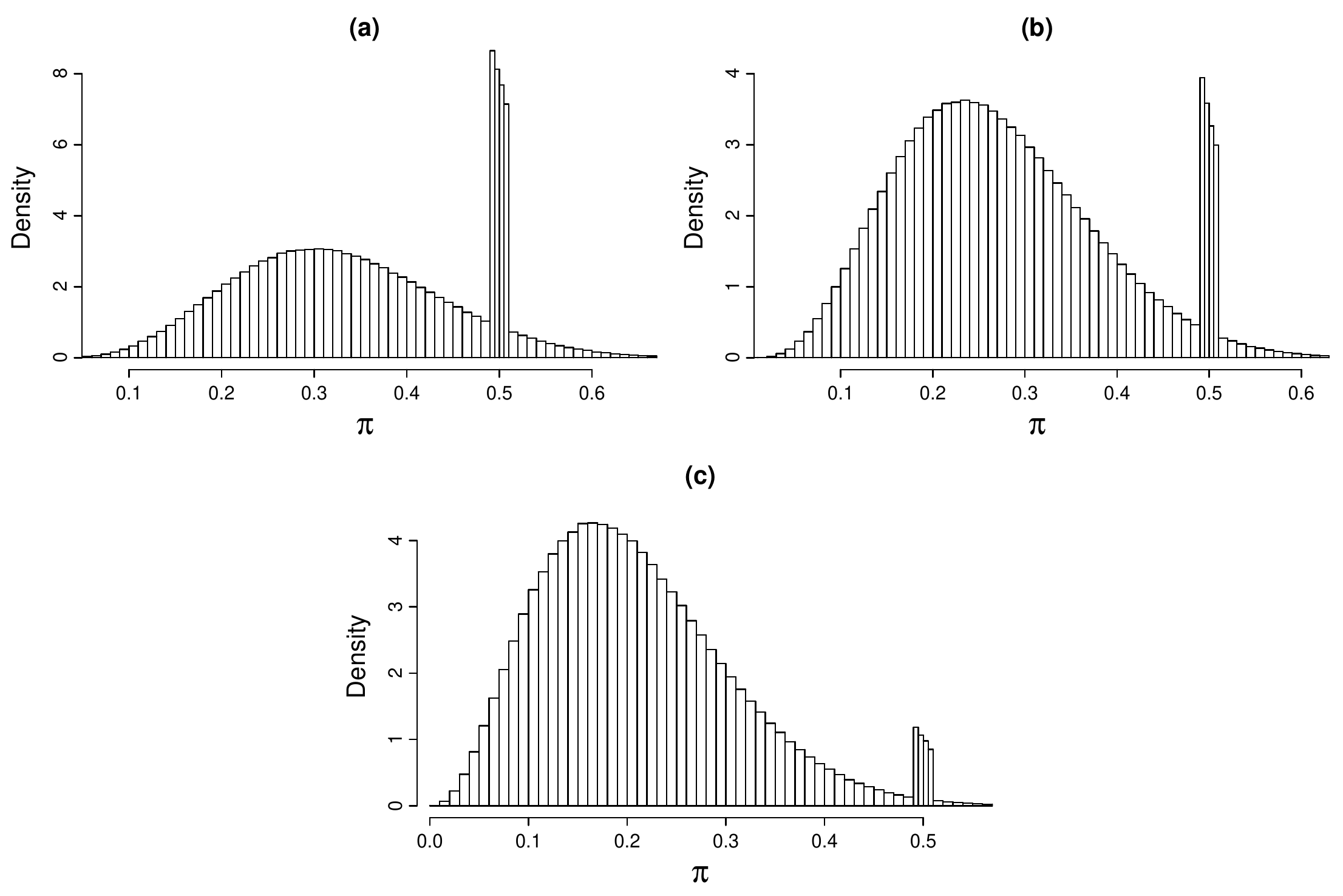}}
\caption{\small{Histograms representing post-data densities of a binomial
proportion $\pi$ for three different observed counts $x$ (namely 5, 4 and 3) in the case where
$n=16$, $\varepsilon=0.01$ and the prior probability that $\pi \in [0.49,0.51]$ is equal to 0.3}}
\end{center}
\end{figure}

It can be seen from Figure~4(a) that when the sample proportion $x/n$ is 0.3125, a substantial
amount of the probability mass of the post-data density $\tilde{p}(\pi\,|\,x)$ can be found in the
interval $[0.49,0.51]$, but from Figures 4(b) and 4(c) it can be seen that the probability mass of
this density that lies in this interval decreases sharply as the sample proportion $x/n$ moves
further away from 0.5. To give some more detail here, the post-data probability $\tilde{p}(\pi \in
[0.49,0.51]\,|\,x)$ is equal to 0.1580, 0.0689 and \pagebreak 0.0204 when the sample count $x$ is
equal to 5, 4 and 3, respectively.

\vspace{3ex}
\section{A more sophisticated method}
\label{sec7}

Although the methodology outlined in Sections~\ref{sec1} to~\ref{sec5} allows us to adequately
determine the post-data probability of $\theta_j$ lying inside the narrow interval
$[\theta_{j0}, \theta_{j1}]$, i.e.\ the probability $\tilde{p}(\theta_j \in [\theta_{j0},
\theta_{j1}]\,|\,\theta_{-j},x)$, and also the post-data density of $\theta_j$ conditional on
$\theta_j$ lying outside the interval $[\theta_{j0}, \theta_{j1}]$, i.e.\ the fiducial density
$f(\theta_j\,|\,\theta_j \notin [\theta_{j0}, \theta_{j1}],\theta_{-j},x)$, it does have two slight
disadvantages.
First, it will generally be the case that the post-data density
$\tilde{p}(\theta_j\,|\,\theta_{-j},x)$ defined by equation~(\ref{equ11}) will have two large and
unrealistic discontinuities at the points $\theta_j=\theta_{j0}$ and $\theta_j=\theta_{j1}$ of the
type that can be clearly seen in Figure~2 and Figures~4(a) to~4(c).
Second, the methodology in question does not take into account that usually there will be more
pre-data belief that $\theta_j$ lies towards the centre of the interval $[\theta_{j0},
\theta_{j1}]$ rather than towards its lower and upper limits.
With the aim of remedying these two drawbacks, we now will put forward a small modification to the
methodology detailed in Sections~\ref{sec2} and~\ref{sec5}.

In particular, let us replace the simple GPD function for $\theta_j$ in the case where
$\theta_j=\theta_{jA}$, i.e.\ where $\theta_j \in [\theta_{j0}, \theta_{j1}]$, that was specified
in equation~(\ref{equ5}) by the following more sophisticated GPD function for $\theta_j$:
\vspace{1ex}
\begin{equation}
\label{equ16}
\oline{\omega}_G (\theta_j) = \left\{
\begin{array}{ll}
1 + \tau h(\theta_j)\, & \mbox{if $\theta_j \in [\theta_{j0}, \theta_{j1}]$}\\[1.25ex]
0 & \mbox{otherwise}
\end{array}
\right.
\vspace{1ex}
\end{equation}
where $\tau \geq 0$ is a given constant and $h(\theta_j)$ is a continuous unimodal density function
on the interval $[\theta_{j0}, \theta_{j1}]$ that is equal to zero at the limits of this interval.
Although, on the basis of this GPD function, we can not determine the fiducial density
$f(\theta_{j}\,|\,\theta_{jA},\theta_{-j},x)$, i.e.\ the density
$f(\theta_j\,|\,\theta_j \in [\theta_{j0}, \theta_{j1}],\theta_{-j},x)$, using Principle~2
summarised in Section~\ref{sec3}, Principle~1 and the extension to Principle~2 discussed in this
earlier section are nevertheless available for this purpose.
If Principle~1 can be applied to the present case, then using this principle would imply that the
fiducial density $f(\theta_{j}\,|\,\theta_{jA},\theta_{-j},x)$ would, in general, be derived by
using the weak fiducial argument.
Furthermore, if Principle~1 or the extension to Principle~2 being referred to can be applied, then
the fiducial density of $\theta_j$ in question would again be defined by the general expression for
the fiducial density $f(\theta_j\,|\,\theta_{-j},x)$ given in equation~(\ref{equ7}).
After determining the fiducial density $f(\theta_{j}\,|\,\theta_{jA},\theta_{-j},x)$ that
corresponds to the GPD function of current interest, the post-data density of $\theta_j$ over the
whole of the real line, i.e.\ the density $\tilde{p}(\theta_j\,|\,\theta_{-j},x)$, can then be
derived in the usual way by using the expression in equation~(\ref{equ11}).

However, there is a critical final issue that needs to be resolved, which is how the value of the
constant $\tau$ in equation~(\ref{equ16}) is chosen. In this regard, let us make the assumption
that the value of $\tau$ is chosen in a way that implies that the post-data \pagebreak density
$\tilde{p}(\theta_j\,|\,\theta_{-j},x)$ will be continuous in the vicinity of the points
$\theta_j=\theta_{j0}$ and $\theta_j=\theta_{j1}$.
In general, a unique value of $\tau$ will exist that satisfies this condition.
Therefore, taking into account also the general form of the proposed GPD function of $\theta_j$
given in equation~(\ref{equ16}), the slight drawbacks can be avoided that were identified at the
start of this section as being associated with the methodology outlined in Sections~\ref{sec2}
and~\ref{sec5}.

Furthermore, there are two reasons why placing the condition in question on the choice of the
constant $\tau$ can be viewed as not placing a substantial restriction on the way we are allowed to
express our pre-data knowledge about the parameter $\theta_j$.
First, since breaking this condition will generally lead to the post-data density of $\theta_j$
having undesirable discontinuities, it can be regarded as being a useful guideline in choosing a
suitable GPD function for $\theta_j$ in the case where it is assumed that $\theta_j$ lies in the
interval $[\theta_{j0}, \theta_{j1}]$.
Second, any detrimental effect caused by obliging the value of $\tau$ to obey the condition under
discussion may not be that apparent given the great deal of imprecision there will usually be in
the specification of the GPD function of $\theta_j$ in the case in question.

\vspace{3ex}
\section{Third example: Revisiting the simple normal case}
\label{sec10}

To give an example of the application of the method proposed in the previous section, let us return
to the problem of making inferences about the mean $\mu$ of a normal distribution that was last
considered in Section~\ref{sec6}.

We will assume that the density function $h(\theta_j)$ that appears in equation~(\ref{equ16}),
i.e.\ the density $h(\mu)$ in the present case, is defined by:
\vspace{-0.5ex}
\begin{equation}
\label{equ17}
\mu \sim \mbox{Beta}\hspace{0.1em} (4,4,-\varepsilon,\varepsilon)
\vspace{-0.5ex}
\end{equation}
i.e.\ it is a beta density function for $\mu$ on the interval $[-\varepsilon, \varepsilon]$ with
both its shape pa\-ram\-e\-ters equal to 4. This density function clearly satisfies the conditions
that were placed on the function $h(\theta_j)$ in the last section.
Observe that choosing the GPD function of $\theta_j$ to be as specified by equation~(\ref{equ16})
implies that the fiducial density $f(\theta_{j}\,|\,\theta_{jA},\theta_{-j},x)$ will be derived in
the current example under Principle~1 described in Section~\ref{sec3} by using the weak fiducial
argument.
As a result, it turns out that this fiducial density, which in the present case is the fiducial
density $f(\mu\,|\,\mu \in [-\varepsilon, \varepsilon],x)$, is \pagebreak defined by:
\begin{equation}
\label{equ20}
f(\mu\,|\,\mu \in [-\varepsilon, \varepsilon],x) = \mathtt{C}_5
(1 + \tau h(\mu)) \phi ((\mu-\bar{x})\sqrt{n}/\sigma)
\end{equation}
where $\mathtt{C}_5$ is a normalising constant, $\phi$ is the standard normal density function and
$h(\mu)$ is as defined in equation~(\ref{equ17}).
On the basis of this fiducial density, the post-data density of $\mu$ over the whole of the real
line, i.e.\ the density $\tilde{p}(\mu\,|\,x)$, can be derived in the usual way by using the
expression in equation~(\ref{equ11}).

To illustrate the use of the calculations just referred to, Figure~5 shows plots of the post-data
density $\tilde{p}(\mu\,|\,x)$ of present interest for various values of the observed mean
$\bar{x}$. As was the case in the construction of Figures~1 to~3, it has been assumed that the
variance $\sigma^2$ is equal to the sample size $n$. However, unlike what was the case in the
construction of the density functions in Figure~2, it has been assumed that $\varepsilon=0.2$
instead of $\varepsilon=0.1$, and as a result, to maintain a useful comparison with this earlier
figure, it has been assumed that the prior probability $p(\mu \in [-\varepsilon,\varepsilon])$ is
equal to 0.33 rather than 0.30. Finally, as was the case in Figure~2, the three post-data densities
of $\mu$ plotted in Figure~5 correspond to the observed mean $\bar{x}$ being equal to 1.7, 2.1 and
2.5. In particular, the short-dashed, solid and long-dashed curves in this figure correspond to
$\bar{x}$ being equal to 1.7, 2.1 and 2.5, respectively. (The line types used have been switched
around in comparison to Figure~2 to improve visualisation.)

\begin{figure}[t]
\begin{center}
\includegraphics[width=5.5in]{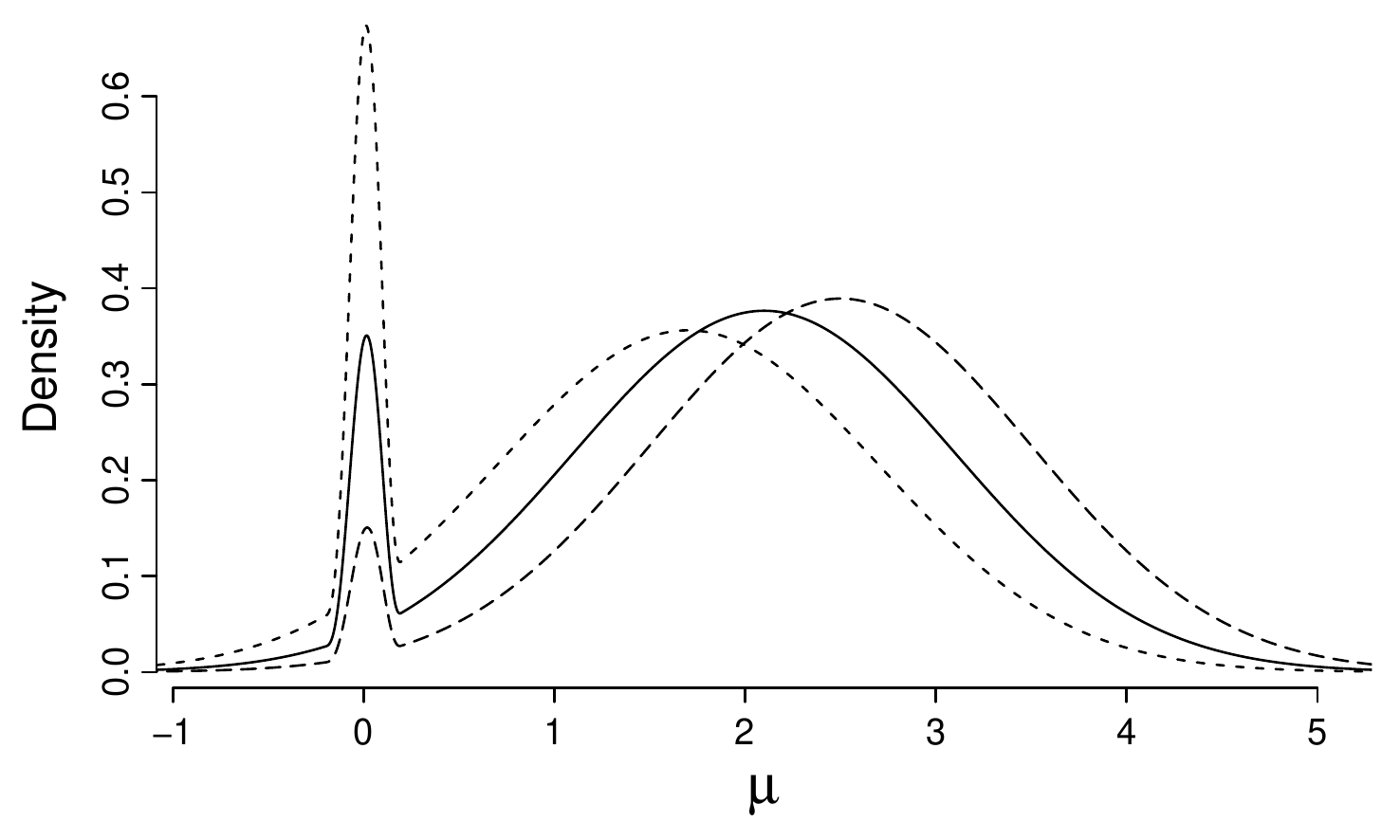}
\caption{\small{Post-data densities of a normal mean $\mu$ for three different values of the sample
mean $\bar{x}$ (namely 1.7, 2.1 and 2.5) in the case where $\varepsilon=0.2$ and the prior
probability that $\mu \in [-0.2,0.2]$ is equal to 0.33}}
\end{center}
\end{figure}

By comparing Figure~5 with Figure~2, it can be appreciated that the application of the method that
was outlined in the previous section will generally allow us to obtain a post-data density
$\tilde{p}(\mu\,|\,x)$ that more realistically represents our post-data knowledge about $\mu$ for
values of $\mu$ both inside and at the limits of the interval $[-\varepsilon, \varepsilon]$ than is
possible using the original method for deriving the post-data density in question that was outlined
in Sections~\ref{sec2} and~\ref{sec5}.

\vspace{3ex}
\section{Extending the methodology to multi-parameter problems}
\label{sec12}

\vspace{0.5ex}
\subsection{General assumptions}
\label{sec9}

It was assumed in Section~\ref{sec1} that the parameter $\theta_j$ is the only unknown parameter in
the sampling model. Let us now assume that all the parameters $\theta_1, \theta_2, \ldots,
\theta_k$ on which the sampling density $g(x\,|\,\theta)$ depends are unknown.

More specifically, we will assume that the subset of parameters $\theta^{(1)} = \{ \theta_1,
\theta_2, \ldots, \theta_m \}$ is such that what would have been believed, before the data were
observed, about each parameter $\theta_{j}$ in this set if all other parameters in the model, i.e.\
$\theta_{-j} = \{ \theta_1, \ldots,   \theta_{j-1},$ $\theta_{j+1},\ldots,\theta_{k} \}$, had been
known, would have satisfied the requirements of the scenario of Definition~2 with $\theta_j$ being
the parameter of interest $\nu$.
Furthermore, it will be assumed that the set of all the remaining parameters in the sampling model,
i.e.\ the set $\theta^{(2)} = \{ \theta_{m+1}, \theta_{m+2}, \ldots, \theta_k \}$, is such that,
before the data were observed, nothing or very little would have been known about each parameter
$\theta_{j}$ in this set over all allowable values of the parameter if all other parameters in the
model, i.e.\ the parameters $\theta_{-j}$, had been known.

It is clear that we can determine the post-data density of any given parameter $\theta_j$ in the
set $\theta^{(1)}$ conditional on all the parameters in the set $\theta_{-j}$ being known by
applying either the method outlined in Sections~\ref{sec2} and~\ref{sec5} or the adjusted version
of this method outlined in Section~\ref{sec7}, meaning that the general definition of this density
function would be as given in equation~(\ref{equ11}).
The set of full conditional post-data densities that result from doing this for each parameter in
the set $\theta^{(1)}$ would therefore be denoted as:
\begin{equation}
\label{equ18}
\tilde{p}(\theta_j\,|\,\theta_{-j},x)\ \ \ \mbox{for $j=1,2,\ldots,m$}
\end{equation}
Also, we can clearly justify specifying the post-data density of any given parameter $\theta_j$ in
the set $\theta^{(2)}$ conditional on all the parameters in the set $\theta_{-j}$ being known as
being the fiducial density $f_{S}(\theta_j\,|\,\theta_{-j},x)$ that was referred to at the end of
Section~\ref{sec3} and which appears in equation~(\ref{equ7}).
Doing this for each parameter in the set $\theta^{(2)}$ would therefore give rise to the following
set of full conditional post-data densities:
\begin{equation}
\label{equ19}
f_{S}(\theta_j\,|\,\theta_{-j},x)\ \ \ \mbox{for $j=m\hspace{-0.05em}+\hspace{-0.05em}1,
m\hspace{-0.05em}+\hspace{-0.05em}2,\ldots,k$}
\end{equation}

\vspace{3ex}
\subsection{Post-data densities of various parameters}
\label{sec11}

If the complete set of full conditional densities of the parameters $\theta_1, \theta_2, \ldots,
\theta_k$ that results from combining the sets of full conditional densities in
equations~(\ref{equ18}) and~(\ref{equ19}) determine a unique joint density for these parameters,
then this density function will be defined as being the post-data density function of $\theta_1,
\theta_2, \ldots, \theta_k$.
However, this complete set of full conditional densities may not be consistent with any joint
density of the parameters concerned, i.e.\ these full conditional densities may be incompatible
among themselves.

To check whether the full conditional densities of $\theta_1, \theta_2, \ldots, \theta_k$ being
referred to are compatible, it may be possible to use an analytical method. An example of an
analytical method that could be used to try to achieve this goal was outlined in relation to full
conditional densities of a similar type in Bowater~(2018).

By contrast, in situations that will undoubtedly often arise where it is not easy to establish
whether or not the full conditional densities defined by equations~(\ref{equ18}) and~(\ref{equ19})
are compatible, let us imagine that we make the pessimistic assumption that they are in fact
incompatible.
Nevertheless, even though these full conditional densities could be incompatible, they could be
reasonably assumed to represent the best information that is available for constructing a joint
density function for the parameters $\theta = \{\theta_1, \theta_2, \ldots, \theta_k\}$ that most
accurately represents what is known about these parameters after the data have been observed, i.e.\
constructing, what could be referred to as, the most suitable post-data density for these
parameters.
Therefore, it would seem appropriate to try to find the joint density of the parameters $\theta$
that has full conditional densities that most closely approximate those given in
equations~(\ref{equ18}) and~(\ref{equ19}).

To achieve this objective, let us focus attention on the use of a method that was advocated in a
similar context in Bowater~(2018), in particular the method that simply consists in making the
assumption that the joint density of the parameters $\theta$ that most closely corresponds to the
full conditional densities in equations~(\ref{equ18}) and~(\ref{equ19}) is equal to the limiting
density function of a Gibbs sampling algorithm (Geman and Geman~1984, Gelfand and Smith~1990) that
is based on these conditional densities with some given fixed or random scanning order of the
parameters in question.
Under a fixed scanning order of the model parameters, we will define a single transition of this
type of algorithm as being one that results from randomly drawing a value (only once) from each of
the full conditional densities in equations~(\ref{equ18}) and~(\ref{equ19}) according to some given
fixed ordering of these densities, replacing each time the previous value of the parameter
concerned by the value that is generated.
Let us clarify that it is being assumed that only the set of values for the parameters $\theta$
that are obtained on completing a transition of this kind are recorded as being a newly generated
sample, i.e.\ the intermediate sets of parameter values that are used in the process of making such
a transition do not form part of the output of the algorithm.

To measure how close the full conditional densities of the limiting density function of the general
type of Gibbs sampler being presently considered are to the full conditional densities in
equations~(\ref{equ18}) and~(\ref{equ19}), we can make use of a method that, in relation to its use
in a similar context, was discussed in Bowater~(2018).
To be able to put this method into effect it is required that the Gibbs sampling algorithm that is
based on the full conditional densities in equations~(\ref{equ18}) and~(\ref{equ19}) would be
irreducible, aperiodic and positive recurrent under all possible fixed scanning orders of the
parameters $\theta$.
Assuming that this condition holds, it was explained in Bowater~(2018), how it may be useful to
analyse how the limiting density function of the Gibbs sampler in question varies over a reasonable
number of very distinct fixed scanning orders of the parameters concerned, remembering that each of
these scanning orders has to be implemented in the way that was just specified.
In particular, it was concluded that if within such an analysis, the variation of this limiting
density with respect to the scanning order of the parameters $\theta$ can be classified as small,
negligible or undetectable, then this should give us reassurance that the full conditional
densities in equations~(\ref{equ18}) and~(\ref{equ19}) are, respectively according to such
classifications, close, very close or at least very close, to the full conditional densities of the
limiting density of a Gibbs sampler of the type that is of main interest, i.e.\ a Gibbs sampler
that is based on any given fixed or random scanning order of the parameters concerned.
See Bowater~(2018) for the line of reasoning that justifies this conclusion.

In trying to choose the scanning order of the parameters $\theta$ such that the type of Gibbs
sampler under discussion has a limiting density function that corresponds to a set of full
conditional densities that most accurately approximate the density functions in
equations~(\ref{equ18}) and~(\ref{equ19}), we should always take into account the precise context
of the problem of inference being analysed.
Nevertheless, a good general choice for this scanning order could arguably be, what will be
referred to as, a uniform random scanning order. Under this type of scanning order, a transition of
the Gibbs sampling algorithm in question will be defined as being one that results from generating
a value from one of the full conditional densities in equations~(\ref{equ18}) and~(\ref{equ19})
that is chosen at random, with the same probability of\hspace{0.1em} $1/k$\hspace{0.1em} being
given to any one of these densities being selected, and then treating the generated value as the
updated value of the parameter concerned.

It is clear that being able to obtain a large random sample from a suitable post-data density of
the parameters $\theta_1, \theta_2, \ldots, \theta_k$ using a Gibbs sampler in the way that has
been described in the present section will usually allow us to obtain good approximations to
expected values of given functions of these parameters over the post-data density concerned, and
thereby allow us to make useful and sensible inferences about the parameters $\theta_1, \theta_2,
\ldots, \theta_k$ on the basis of the data set $x$ of interest.

\vspace{3ex}
\section{Fourth example: Normal case with variance unknown}
\label{sec14}

To demonstrate the application of the approach to multi-parameter problems that has just been
outlined, let us return to the example that was the focus of our attention in Section~\ref{sec8},
however let us now assume that the difference in the performance of the two drugs of interest is
measured by the difference in the concentration of a certain chemical (e.g.\ cholesterol) in the
blood, in particular the level observed for drug A minus the level observed for drug B, rather than
the preferences of the patients concerned.
The set of these differences for all of the $n$ patients in the sample will be the data set $x$.
This example therefore also has a good deal in common with the example that was considered in
Section~\ref{sec6}. Moreover, similar to this earlier example, it will be assumed that each value
$x_i$ in the data set $x$ follows a normal distribution with an unknown mean $\mu$, however, by
contrast to this previous example, the standard deviation $\sigma$ of this distribution will now
also be treated as being unknown.

For the same type of reason to that used in Example~1.2 in the Introduction, let us in addition
suppose that, for any given value of $\sigma$, the scenario of Definition~2 would apply in relation
to the parameter $\mu$ with the special hypothesis being the hypothesis that $\mu$ lies in the
narrow interval $[-\varepsilon, \varepsilon]$.
On the other hand, it will be assumed, as could often be done in practice, that nothing or very
little would have been known about the standard deviation $\sigma$ before the data were observed
given any value for $\mu$. Therefore, in terms of the notation of Section~\ref{sec9}, the set of
parameters $\theta^{(1)}$ will only contain $\mu$, and the set $\theta^{(2)}$ will only
contain~$\sigma$.

Taking into account our pre-data knowledge about $\mu$ conditional on any given value of $\sigma$,
we may choose to define the post-data density of $\mu$ conditional on $\sigma$ using either the
method outlined in Sections~\ref{sec2} and~\ref{sec5} or the adjusted version of this method
outlined in Section~\ref{sec7}. Let us choose the latter option. In particular, we will assume that
this post-data density is specified in the same way as the post-data density $\tilde{p}(\mu\,|\,x)$
was specified in the example outlined in Section~\ref{sec10}. To clarify, the fiducial density
$f(\mu\,|\,\mu \notin [-\varepsilon, \varepsilon],\sigma,x)$ that is required by
equation~(\ref{equ11}) will be assumed to be a normal density with mean $\bar{x}$ and variance
$\sigma^2/n$ conditioned on $\mu$ lying outside of the interval $[-\varepsilon, \varepsilon]$, and
the fiducial density $f(\mu\,|\,\mu \in [-\varepsilon, \varepsilon],\sigma,x)$, which also needs to
enter into equation~(\ref{equ11}), will be assumed to be defined as the fiducial density of $\mu$
was defined in equation~(\ref{equ20}). It can be seen that we are not making any distinction here
between deriving a post-data density of $\mu$ in a situation where $\sigma$ is known and deriving a
post-data density of $\mu$ conditional on $\sigma$ taking a given value.

Furthermore, given our lack of any substantial pre-data knowledge about $\sigma$ conditional on any
particular value of $\mu$, it would seem natural to derive the fiducial density of $\sigma$
conditional on $\mu$ on the basis of a GPD function of $\sigma$ that is defined as follows:
$\omega_{G} (\sigma) = a$ if $\sigma \geq 0$ and zero otherwise, where $a>0$. Observe that since
the variance estimator $\bm\hat{\sigma}^2 = (1/n)\sum_{i=1}^{n} (x_i - \mu)^2$ is a sufficient
statistic for $\sigma$ if $\mu$ is known, it can therefore be assumed to be the fiducial statistic
$Q(x)$ in deriving the fiducial density of $\sigma$ in question. Based on this assumption, the
function $\varphi$ that forms part of the data generating algo-\linebreak rithm described in
Section~\ref{sec3} can be expressed as:
\[
\varphi(\Gamma,\sigma^2)=(\sigma^2/n)\Gamma
\]
where the primary r.v.\ $\Gamma$ has a $\chi^2$ distribution with $n$ degrees of freedom, which
means that it is being assumed that the fiducial statistic $\bm\hat{\sigma}^2$ is determined by
setting $\Gamma$ equal to its already generated value $\gamma$ in the formula:
$\bm\hat{\sigma}^2 = \varphi(\Gamma,\sigma^2)$.

Taking into account the way the GPD function of $\sigma$ was just specified, we may then derive the
fiducial density of $\sigma^2$ conditional on $\mu$ under Principle~1 described in
Section~\ref{sec3} by using the strong fiducial argument. As a result, it turns out that this
fiducial density, which to be consistent with the notation used earlier will be denoted as the
density $f_{S}(\sigma^2\,|\,\mu,x)$, can be expressed as:
\begin{equation}
\label{equ21}
\sigma^2\,|\,\mu,x \sim \mbox{Inv-Gamma}\, (\alpha = n/2,\hspace{0.1em}
\beta = n\bm\hat{\sigma}^2/2)
\end{equation}
i.e.\ it is an inverse gamma density function with shape parameter $\alpha$ equal to $n/2$ and
scale parameter $\beta$ equal to $n\bm\hat{\sigma}^2/2$.
In accordance with what was discussed in Section~\ref{sec9}, this density function will therefore
be regarded as being the post-data density of $\sigma^2$ conditional on $\mu$ in the example being
considered.

To develop the analysis of this example further, Figure~6 shows some results from running a Gibbs
sampler on the basis of the full conditional post-data densities of $\mu$ and $\sigma$ that have
just been defined, i.e.\ the densities $\tilde{p}(\mu\,|\,\sigma,x)$ and
$f_{S}(\sigma\, |\,\mu, x)$, with a uniform random scanning order of the parameters $\mu$ and
$\sigma$, as such a scanning order was defined in Section~\ref{sec11}.
In particular, the histograms in Figures~6(a) and~6(b) represent the distributions of the values of
$\mu$ and $\sigma$, respectively, over a single run of five million samples of these parameters
generated by the Gibbs sampler after a preceding run of one thousand samples, which were classified
as belonging to its burn-in phase, had been discarded.
This analysis is based on the assumption that the sample size $n$ is equal to 9, the sample mean
$\bar{x}$ is equal to 2.1 and the sample standard deviation $s$ is equal to 3, meaning of course
that the usual estimator of the standard error of the sample mean, i.e.\ $s/\sqrt{n}$, is equal to
one. Also, similar to the example discussed in Section~\ref{sec10}, it was assumed, in specifying
the post-data density $\tilde{p}(\mu\,|\,\sigma,x)$, that $\varepsilon=0.2$ and that the prior
probability $p_0(\theta_{jA})$ in equation~(\ref{equ10}), which in the present case is the
\pagebreak prior probability that $\mu \in [-\varepsilon,\varepsilon]$, is equal to 0.33.

\begin{figure}[t]
\begin{center}
\makebox[\textwidth]{\includegraphics[width=7.7in]{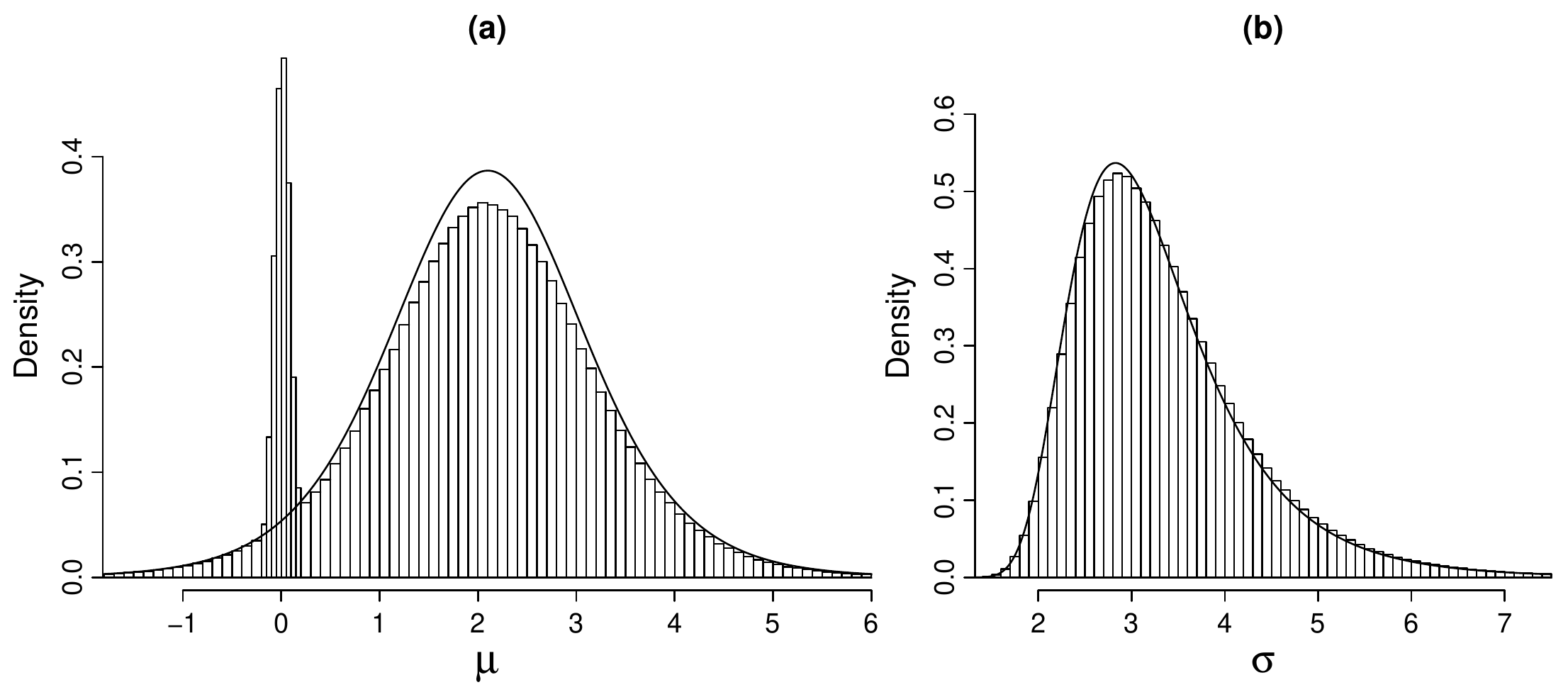}}
\caption{\small{Marginal post-data densities of the mean $\mu$ and the standard deviation $\sigma$
of a normal distribution in the case where the sample mean $\bar{x}$ equals 2.1, the sample
standard deviation $s$ equals 3 and the sample size $n$ equals 9}}
\end{center}
\end{figure}

In accordance with conventional recommendations for evaluating the convergence of Monte Carlo
Markov chains outlined, for example, in Gelman and Rubin~(1992) and Brooks and Roberts~(1998), an
additional analysis was carried out in which the Gibbs sampler was run various times from different
starting points and the output of these runs was carefully assessed for convergence using suitable
diagnostics. This analysis provided no evidence to indicate that the sampler does not have a
limiting distribution, and showed, at the same time, that it would appear to generally converge
quickly to this distribution.

Furthermore, the Gibbs sampling algorithm was run separately with each of the two possible fixed
scanning orders of the parameters $\mu$ and $\sigma$, i.e.\ the one in which $\mu$ is updated first
and then $\sigma$ is updated, and the one that has the reverse order, in accordance with how a
single transition of such an algorithm was defined in Section~\ref{sec11}, i.e.\ single transitions
of the algorithm incorporated updates of both parameters.
In doing this, the sample correlations between $\mu$ and $\sigma$ were 0.075 and 0.120 over the
runs of the sampler that were based on the first scanning order just mentioned and the reverse
scanning order, respectively, after excluding the burn-in phase of the sampler, and the difference
between these two correlations was statistically significant (meaning simply that the difference
was beyond random chance).
Taking into account results referred to in Section~\ref{sec11}, it can therefore be concluded that
the two full conditional densities on which the Gibbs sampler is based, i.e.\ the post-data
densities $\tilde{p}(\mu\,|\,\sigma,x)$ and $f_{S}(\sigma\, |\,\mu, x)$, are incompatible.
However, as discussed in Section~\ref{sec11}, even though these post-data conditional densities are
incompatible they could be reasonably assumed to represent the best information that is available
for constructing a joint density function for the parameters $\mu$ and $\sigma$ that most
accurately represents what is known about these parameters after the data have been observed.

To investigate further the issue of how close the full conditional densities of the limiting
density function of the type of Gibbs sampler being considered are to the full conditional
densities\hspace{0.05em} $\tilde{p}(\mu\,|\,\sigma,x)$ and $f_{S}(\sigma\, |\,\mu, x)$, another
supplementary study was conducted in which the values of $\mu$ and $\sigma$ generated by the Gibbs
sampler in question were used to approximate the former type of full conditional densities of $\mu$
and $\sigma$.
The results of this study showed that, independent of the scanning order of the parameters
concerned, the full conditional densities of the limiting density function under discussion may be
arguably regarded as being adequate approximations to the full conditional densities
$\tilde{p}(\mu\,|\,\sigma,x)$ and $f_{S}(\sigma\, |\,\mu, x)$. Also, in cases such as the present
example where the sample size $n$ is small, it can be argued that we should be a little less strict
in how we judge the adequacy of the approximations in question.
To clarify, the results of using three different types of scanning order of $\mu$ and $\sigma$ were
analysed in this study and these, in particular, were the uniform random scanning order of these
parameters and the two fixed scanning orders of $\mu$ and $\sigma$ that were just described.

Let us point out that, although the limiting density function of the Gibbs sampler being considered
is affected to some extent by what choice is made for the scanning order of $\mu$ and $\sigma$,
the marginal densities of $\mu$ and $\sigma$ over this joint limiting density of $\mu$ and $\sigma$
will be the same whatever scanning order of $\mu$ and $\sigma$ is used.
If the aim is to only obtain marginal post-data densities of $\mu$ and $\sigma$, then this property
is clearly convenient as it means we can avoid needing to explicitly determine the scanning order
of $\mu$ and $\sigma$ that corresponds to the limiting density of the Gibbs sampler having full
conditional densities that, in some way, optimally approximate the conditional post-data densities
$\tilde{p}(\mu\,|\,\sigma,x)$ and $f_{S}(\sigma\, |\,\mu, x)$.

With regard to analysing the data set $x$ of current interest, the curves overlaid on the
histograms in Figures~6(a) and~6(b) are plots of the marginal fiducial densities of the parameters
$\mu$ and $\sigma$, respectively, in the case where the joint fiducial density of these parameters
is defined by the conditional fiducial density $f_{S}(\sigma\,|\,\mu, x)$ that is specified by
equation~(\ref{equ21}) and by the fiducial density of $\mu$ conditional on $\sigma$ that, similar
to how we derived the density $f_{S}(\sigma\,|\,\mu, x)$, is derived by using the strong fiducial
argument, meaning that this conditional density of $\mu$ is defined by the following expression:
\vspace{-0.5ex}
\[
\mu\,|\,\sigma,x \sim \mbox{N}(\bar{x}, \sigma^2/n)
\vspace{-0.5ex}
\]
Observe that this latter conditional fiducial density, which we will naturally denote as the
density $f_{S}(\mu\,|\,\sigma,x)$, is compatible with the conditional density
$f_{S}(\sigma\,|\,\mu, x)$ and that the joint fiducial density of $\mu$ and $\sigma$ that they
directly define is unique. To clarify, it would be reasonable to regard the fiducial density
$f_{S}(\mu\,|\,\sigma,x)$ as representing our post-data knowledge about $\mu$ conditional on
$\sigma$ if nothing or very little was known about $\mu$ before the data were observed. Here we are
therefore treating both the parameters $\mu$ and $\sigma$ as belonging to the set $\theta^{(2)}$
referred to in Section~\ref{sec9}.

The joint fiducial density of $\mu$ and $\sigma$ just referred to, which we will naturally denote
as the density $f_S(\mu, \sigma\,|\,x)$, can be expressed as follows:
\begin{equation}
\label{equ26}
f_S(\mu, \sigma\,|\,x) = \mathtt{C}_6 ( 1/\sigma^2)^{(n/2)+1}
\exp\{-(1/2\sigma^2)((n-1)s^2 + n (\bar{x} - \mu)^2 )\}
\end{equation}
where $\mathtt{C}_6$ is a normalising constant. Given this is the joint fiducial density of $\mu$
and $\sigma$, the marginal fiducial density of $\mu$, which we will naturally denote as the density
$f_S(\mu\,|\,x)$, and which, for the data $x$ of current interest, is represented by the curve in
Figure~6(a), can be expressed as follows:
\vspace{-0.5ex}
\begin{equation}
\label{equ29}
\mu\,|\,x \sim \mbox{Non-standardised}\ t_{n-1}(\bar{x}, s/\sqrt{n}\hspace{0.1em})
\vspace{-0.5ex}
\end{equation}
i.e.\ it is a non-standardised Student $t$ density function with $n-1$ degrees of freedom, location
parameter equal to $\bar{x}$ and scaling parameter equal to $s/\sqrt{n}$. \pagebreak
Finally, the corresponding marginal fiducial density of $\sigma$, which we will naturally denote as
the density $f_S(\sigma\,|\,x)$, and which, for the data $x$ of current interest, is represented by
the curve in Figure~6(b), is given by the expression:
\begin{equation}
\label{equ32}
\sigma^2\,|\,x \sim \mbox{Inv-Gamma}\, ((n-1)/2,\hspace{0.1em} (n-1)s^2/2)
\end{equation}

Observe that the accumulation of probability mass around the value of zero in the marginal
post-data density of $\mu$ that is represented by the histogram in Figure~6(a), and the fact that
the upper tail of the marginal post-data density of $\sigma$ that is represented by the histogram
in Figure~6(b) tapers down to zero slightly more slowly than the marginal fiducial density for
$\sigma$ that is represented by the curve in Figure~6(b) are both consequences that would be
expected of the strong pre-data opinion about $\mu$ that was incorporated into the derivation of
the conditional post-data density $\tilde{p}(\mu\,|\,\sigma,x)$.

\vspace{3ex}
\section{Direct extension of the univariate method to the multivariate case}
\label{sec15}

Although it was demonstrated in the previous section how the methodology of Section~\ref{sec12} can
be successfully applied to make inferences about the parameters of a normal distribution in the
case where, in terms of the notation of Section~\ref{sec9}, the mean $\mu$ belongs to the set
$\theta^{(1)}$ and the standard deviation $\sigma$ belongs to the set $\theta^{(2)}$, we are
nevertheless left with some motivation to try to find an alternative method that avoids the need to
determine a joint density of $\mu$ and $\sigma$ that has full conditional densities that adequately
approximate incompatible full conditional post-data densities of these parameters.
In this regard, with the aim of making inferences about the parameters $\theta$ of a general
sampling density $g(x\,|\,\theta)$, an alternative method for dealing with the kind of inferential
problem being discussed will be developed in this section that avoids the particular type of
requirement just mentioned and that we may reasonably anticipate will be a useful method in many
situations.

Let us begin by assuming that the parameter $\theta_{j}$ belongs to the set $\theta^{(1)}$ and the
parameters $\theta_{-j}$ belong to the set $\theta^{(2)}$ according to how the sets of parameters
$\theta^{(1)}$ and $\theta^{(2)}$ were defined in Section~\ref{sec9}.
Under this assumption, we will now generalise the second type of Bayesian analogy that was put
forward in Section~\ref{sec4} to the case where all the parameters in the set $\theta = \{\theta_1,
\theta_2, \ldots, \theta_k\}$ are unknown. In particular, let us make an analogy between, on the
one hand, our pre-data uncertainty both about the parameter values in the set $\theta$ and the
values that will make up the data set $x$, and on the other hand, our uncertainty about the outcome
of a physical experiment that will randomly select either a set of values\hspace{0.05em}
$\theta^{A}=\{\theta^{A}_1,\theta^{A}_2,\ldots,\theta^{A}_k\}$ or a set of values\hspace{0.05em}
$\theta^{B}=\{\theta^{B}_1,\theta^{B}_2,\ldots,\theta^{B}_k\}$ for the parameters in the set
$\theta$, and then will generate the data $x$ by substituting these values for the parameters
$\theta$ into the formula of the sampling density $g(x\,|\,\theta)$.
The key detail that establishes this more general type of analogy as being an acceptable one to
make is that it will be assumed that we know that the value for the parameter $\theta_j$ in the set
$\theta^{A}$, i.e.\ the value $\theta^{A}_j$, lies in the interval $[\theta_{j0},\theta_{j1}]$ and
also that the value for this parameter in the set $\theta^{B}$, i.e.\ the value $\theta^{B}_j$,
lies outside of this interval, but apart from this, the values in the sets $\theta^{A}$ and
$\theta^{B}$ will be assumed to be completely unknown to us.

To clarify, in conducting the physical experiment of interest, the joint density of the parameters
$\theta$ and the values in the data set $x$ would be given by:
\vspace{-0.5ex}
\[
p(\theta,x) = p(\theta)g(x\,|\,\theta)
\vspace{-0.5ex}
\]
in which
\vspace{1.5ex}
\begin{equation}
\label{equ23}
p(\theta) = \left\{
\begin{array}{ll}
p_0(\theta^{A}) & \mbox{if $\theta = \theta^{A}$}\\[0.75ex]
p_0(\theta^{B}) = 1 - p_0(\theta^{A})\, & \mbox{if $\theta = \theta^{B}$}\\[0.75ex]
0 & \mbox{otherwise}
\end{array}
\right.
\vspace{2.5ex}
\end{equation}
where $p_0(\theta^{A})$ is a specified probability.
Given our lack of information concerning the parameter values in the sets $\theta^{A}$ and
$\theta^{B}$, the probability $p_0(\theta^{A})$ is in effect our prior probability that the
parameter $\theta_j$ lies in the interval $[\theta_{j0},\theta_{j1}]$, while $p_0(\theta^{B})$ is
in effect our prior probability that $\theta_j$ lies outside of this interval.

Similar to the sampling densities $g(x\,|\,\theta_{jA},\theta_{-j})$ and
$g(x\,|\,\theta_{jB},\theta_{-j})$ referred to in Sections~\ref{sec2} and~\ref{sec5}, the sampling
densities $g(x\,|\,\theta^{A})$ and $g(x\,|\,\theta^{B})$ just referred to are unknown due to there
being a lack of information about the true values of the conditioning variables, which in the
present case are the sets of values $\theta^{A}$ and $\theta^{B}$.
Therefore, similar to what was done in Section~\ref{sec2}, we will choose to estimate the densities
$g(x\,|\,\theta^{A})$ and $g(x\,|\,\theta^{B})$ using the fiducial predictive densities of a future
data set $x'$ that result from separately applying the condition that $\theta = \theta^{A}$ and
that $\theta = \theta^{B}$, respectively, which means that these predictive densities are given by
the expressions:
\vspace{1ex}
\begin{equation}
\label{equ24}
f(x'\,|\,\theta^{A},x) = \int \cdots \int_{\mbox{\footnotesize $\Theta$}^{A}}\hspace{0.1em}
g(x'\,|\,\theta)f(\theta\,|\,\theta^{A},x) d\theta^{A}_1 \ldots d\theta^{A}_k
\end{equation}
and
\vspace{1ex}
\begin{equation}
\label{equ25}
f(x'\,|\,\theta^{B},x) = \int \cdots \int_{\mbox{\footnotesize $\Theta$}^{B}}\hspace{0.1em}
g(x'\,|\,\theta)f(\theta\,|\,\theta^{B},x) d\theta^{B}_1 \ldots d\theta^{B}_k
\vspace{2.5ex}
\end{equation}
respectively, where:\\
i) The functions $f(\theta\,|\,\theta^{A},x)$ and $f(\theta\,|\,\theta^{B},x)$ are joint fiducial
densities of the set of parameters $\theta$ that are determined under the condition that
$\theta=\theta^{A}$ and that $\theta=\theta^{B}$, respectively. In particular, these fiducial
densities are derived by using the methodology outlined in Bowater~(2019b, 2021b) on the basis of
whatever is precisely assumed regarding our pre-data knowledge about the parameters
$\theta$\hspace{0.05em};\\
ii) The domains $\Theta^{A}$ and $\Theta^{B}$ are the spaces of all possible values for the sets
$\theta^{A}$ and $\theta^{B}$, respectively\hspace{0.05em};\\
iii) The future data set $x'$ is again the same size as the observed data set $x$.

Let us now choose to simply point out that by applying the same logic that was used to derive the
post-data probabilities $\tilde{p}(\theta_j = \theta_{jA}\,|\,\theta_{-j},x)$ and
$\tilde{p}(\theta_j = \theta_{jB}\,|\,\theta_{-j},x)$ in equations~(\ref{equ10}) and~(\ref{equ22}),
we may justify that the post-data probabilities of the hypotheses that $\theta = \theta^{A}$ and
that $\theta = \theta^{B}$ are given by the expressions:
\vspace{1ex}
\begin{equation}
\label{equ27}
\tilde{p}(\theta = \theta^{A}\,|\,x) =
\frac{p_0(\theta^{A})f(x' = x\,|\,\theta^{A},x) }
{p_0(\theta^{A})f(x' = x\,|\,\theta^{A},x) +
p_0(\theta^{B})f(x' = x\,|\,\theta^{B},x)}
\end{equation}
and
\vspace{0.5ex}
\[
\tilde{p}(\theta = \theta^{B}\,|\,x) = 1 - \tilde{p}(\theta = \theta^{A}\,|\,x)
\vspace{0.5ex}
\]
respectively, where the notation being used here is the same as used in equations~(\ref{equ23}),
(\ref{equ24}) and~(\ref{equ25}).
Of course, according to the definitions of the sets of values $\theta^{A}$ and $\theta^{B}$ that we
have been using, the probabilities $\tilde{p}(\theta = \theta^{A}\,|\,x)$ and
$\tilde{p}(\theta = \theta^{B}\,|\,x)$ are, in effect, our post-data probabilities that $\theta_j$
lies in the interval $[\theta_{j0},\theta_{j1}]$ and that $\theta_j$ lies outside of this interval,
respectively, and therefore, we may also denote these probabilities as
$\tilde{p}(\theta_j \in [\theta_{j0}, \theta_{j1}]\,|\,x)$ and
$\tilde{p}(\theta_j \notin [\theta_{j0}, \theta_{j1}]\,|\,x)$, respectively.

Furthermore, it is perfectly consistent with the line of reasoning being currently followed to
define the marginal post-data density function of $\theta_j$ over the whole \pagebreak of the real
line in the following way:
\[
\tilde{p}(\theta_j\,|\,x) = f(\theta_j\,|\,\theta^{A},x)\hspace{0.05em}
\tilde{p}(\theta = \theta^{A}\,|\,x) + f(\theta_j\,|\,\theta^{B},x)\hspace{0.05em}
\tilde{p}(\theta = \theta^{B}\,|\,x)
\]
or equivalently in the following way:
\begin{eqnarray}
\label{equ28}
\tilde{p}(\theta_j\,|\,x) & = & f(\theta_{j}\,|\,\theta_j \in
[\theta_{j0}, \theta_{j1}],x)\hspace{0.05em} \tilde{p}(\theta_j \in
[\theta_{j0}, \theta_{j1}]\,|\,x) \nonumber \\
& & + \hspace{0.2em}
f(\theta_j\,|\,\theta_j \notin [\theta_{j0}, \theta_{j1}],x)\hspace{0.05em}
\tilde{p}(\theta_j \notin [\theta_{j0}, \theta_{j1}]\,|\,x)
\end{eqnarray}
\par \vspace{0.5ex} \noindent
where $f(\theta_j\,|\,\theta^{A},x)$ is the marginal density of $\theta_{j}$ over the joint
fiducial density $f(\theta\,|\,\theta^{A},x)$ that was referred to in equation~(\ref{equ24}) and
where $f(\theta_j\,|\,\theta^{B},x)$ is the marginal density of $\theta_{j}$ over the joint
fiducial density $f(\theta\,|\,\theta^{B},x)$ that was referred to in equation~(\ref{equ25}).
Finally, the joint post-data density of the set of parameters $\theta$ may be defined as follows:
\begin{equation}
\label{equ37}
\tilde{p}(\theta\,|\,x) = f(\theta\,|\,\theta^{A},x)\hspace{0.05em}
\tilde{p}(\theta = \theta^{A}\,|\,x) +
f(\theta\,|\,\theta^{B},x)\hspace{0.05em} \tilde{p}(\theta = \theta^{B}\,|\,x)
\end{equation}

Let us now briefly compare the approach that was outlined in Sections~\ref{sec13} and~\ref{sec12}
with the method that has just been put forward. In particular, it can be appreciated that the
method presented in Sections~\ref{sec2} and~\ref{sec5} determines the post-data probability of
$\theta_j$ lying in the interval $[\theta_{j0}, \theta_{j1}]$ given values for the parameters in
the set $\theta_{-j}$ on the basis of the difference in predictions made about a future data set
$x'$ when the fiducial density of $\theta_j$ conditional on the set $\theta_{-j}$ is restricted to
lie in the interval $[\theta_{j0}, \theta_{j1}]$, and when this fiducial density is restricted not
to lie in this interval. To be able to use this earlier approach to obtain a post-data probability
of $\theta_j$ lying in the interval $[\theta_{j0}, \theta_{j1}]$ that is not conditioned on the
parameters $\theta_{-j}$ requires the additional methodology that was put forward in
Section~\ref{sec12}.
On the other hand, the method that has been outlined in the present section determines the
post-data probability of $\theta_j$ lying in the interval $[\theta_{j0}, \theta_{j1}]$ without
conditioning on the parameters $\theta_{-j}$ on the basis of the difference in predictions made
about a future data set $x'$ when the joint fiducial density of $\theta$ is restricted by the
condition that the parameter $\theta_j$ must lie in the interval $[\theta_{j0}, \theta_{j1}]$, and
when this joint fiducial density is restricted by the condition that $\theta_j$ must not lie in
this interval.

As a result, it would appear that this latter method may be particularly useful, in comparison to
the method described in Sections~\ref{sec13} and~\ref{sec12}, when the derivation of the joint
fiducial densities $f(\theta\,|\,\theta^{A},x)$ and $f(\theta\,|\,\theta^{B},x)$ is more
straightforward than trying to derive the joint post-data density of the parameters $\theta$ by
using the approach detailed in Section~\ref{sec12}.
For example, this may well be the case when the full conditional densities in
equations~(\ref{equ18}) and~(\ref{equ19}) are not compatible, but by contrast, the fiducial
densities $f(\theta\,|\,\theta^{A},x)$ and $f(\theta\,|\,\theta^{B},x)$ can be derived using the
methodology outlined in Bowater~(2019b, 2021b) on the basis of sets of full conditional fiducial
densities of the parameters $\theta$ that are indeed compatible.

\vspace{3ex}
\section{Fifth example: Re-analysing the normal case with variance unknown}
\label{sec16}

Let us now apply the method that was put forward in the previous section to the same problem of
inference that was analysed in Section~\ref{sec14}, i.e.\ the problem of making inferences about
the mean $\mu$ and variance $\sigma^2$ of a normal distribution on the basis of a sample $x$
drawn from the distribution concerned.

We will assume that if we found ourselves in the scenario where $\mu$ is conditioned to lie in the
interval $[-\varepsilon, \varepsilon]$ or in the scenario where $\mu$ is conditioned not to lie in
this interval, then we would have had very little or no pre-data knowledge about $\mu$ conditional
on any given value of $\sigma$ and very little or no pre-data knowledge about $\sigma$ conditional
on any given value of $\mu$.
In these two scenarios, it is therefore quite natural that the GPD functions of $\mu$ conditional
on $\sigma$ being known and the GPD functions of $\sigma$ conditional on $\mu$ being known are
specified as being neutral GPD functions (see Section~\ref{sec2} and Bowater~2019b) with the result
that the fiducial density $f(\theta\,|\,\theta^{A},x)$, i.e.\ the density
$f(\mu,\sigma\,|\,\mu \in [-\varepsilon, \varepsilon],x)$ in the present case, and the fiducial
density $f(\theta\,|\,\theta^{B},x)$, i.e.\ the density $f(\mu,\sigma\,|\,\mu \notin [-\varepsilon,
\varepsilon],x)$ in the present case, can be obtained by simply conditioning the fiducial density
$f_{S}(\mu,\sigma\,|\,x)$ defined in equation~(\ref{equ26}) on the requirement that $\mu \in
[-\varepsilon, \varepsilon]$ and that $\mu \notin [-\varepsilon, \varepsilon]$, respectively.
In order to be able to use equation~(\ref{equ27}) to determine the post-data probability
$\tilde{p}(\theta = \theta^{A}\,|\,x)$, i.e.\ the probability $\tilde{p}(\mu \in [-\varepsilon,
\varepsilon]\,|\,x)$ in the present case, we are hence left with only the task of needing to assign
a value to the probability $p_0(\theta^{A})$, i.e.\ the task of assigning a prior probability to
the event of $\mu$ lying in the interval $[-\varepsilon, \varepsilon]$.

To illustrate the use of the calculation that appears in equation~(\ref{equ27}) in the present
example, Figure~7 shows various plots of the post-data probability $\tilde{p}(\mu \in
[-\varepsilon, \varepsilon]\,|\,x)$ against the observed mean $\bar{x}$. Similar to the example
discussed in Section~\ref{sec14}, this figure is based on the assumption that the sample size $n$
is equal to 9 and the sample standard deviation $s$ is equal to 3, meaning of course that the usual
estimator of the standard error of the sample mean, i.e.\ $s/\sqrt{n}$, is equal to one.
As was the case in Figure~1, the short-dashed, long-dashed and solid curves that form the three
upper curves in Figure~7 correspond to the scenario where the prior probability
$p(\mu \in [-\varepsilon,\varepsilon])$ is equal to 0.5, while the three lower curves in this
figure correspond to the scenario where the prior probability
$p(\mu \in [-\varepsilon,\varepsilon])$ is equal to 0.3. Also similar to Figure~1, the
short-dashed, long-dashed and solid curves in Figure~7 correspond to the cases where
$\varepsilon = 0.5$, $\varepsilon = 0.25$ and $\varepsilon = 0$, respectively.

\begin{figure}[t]
\begin{center}
\includegraphics[width=5in]{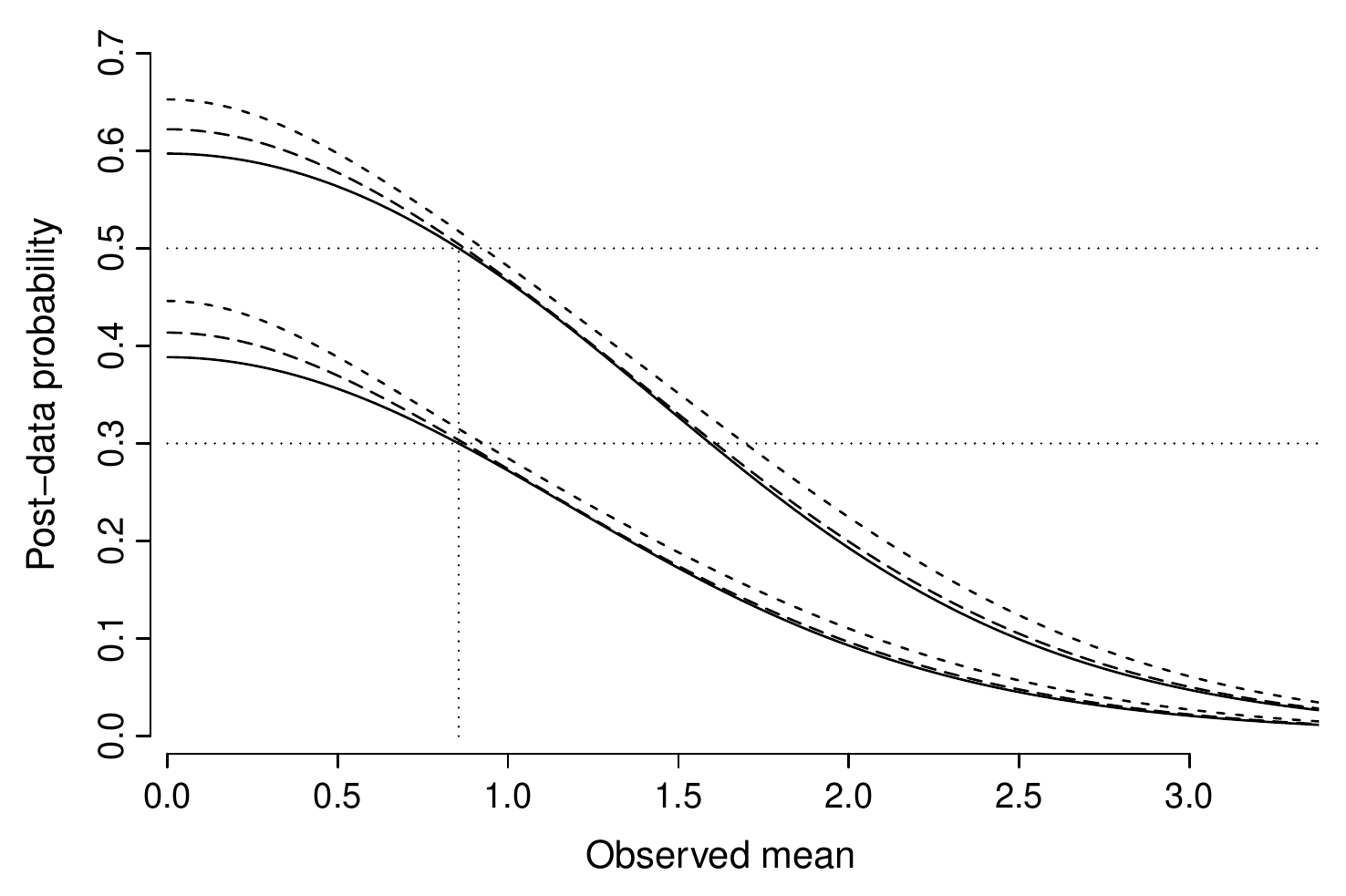}
\caption{\small{Post-data probability of a normal mean $\mu$ lying in the interval
$[-\varepsilon, \varepsilon]$ against the sample mean $\bar{x}$ in the case where the standard
deviation $\sigma$ is unknown. Plotted curves correspond to three values of $\varepsilon$ (namely
0, 0.25 and 0.5) and two values for the prior probability of $\mu \in [-\varepsilon, \varepsilon]$
(namely 0.3 and 0.5)}}
\end{center}
\end{figure}

It can be seen from the curves in Figure~7 that, in any of the scenarios of interest, to be able to
reduce the post-data probability $\tilde{p}(\mu \in [-\varepsilon, \varepsilon]\,|\,x)$ down to any
given low level generally requires larger values of the sample mean $\bar{x}$ than was the case
under the assumptions on which Figure~1 is based. This is of course not at all surprising given the
uncertainty there is in the present example about the true value of the standard deviation
$\sigma$. In the case where $\varepsilon=0$, it is also of interest to note that for the
three\linebreak values of the sample mean $\bar{x}$ that correspond to the two-sided P value of the
Student $t$ test of the null hypothesis that $\mu = 0$ being 0.05, 0.01 and 0.001, the post-data
prob\-a\-bil\-ity $\tilde{p}(\mu = 0 \,|\,x)$ would be 0.1301, 0.0277 and 0.0024, respectively, if
the prior probability that $\mu = 0$ was 0.5, and would be 0.0602, 0.0120 and 0.0010, respectively,
if this prior probability was~0.3.

To obtain the marginal post-data density of $\mu$, i.e.\ the density $\tilde{p}(\mu\,|\,x)$, we may
use the expression in equation~(\ref{equ28}). In the present example, this general expression may
be converted into the following more specific formula:
\begin{eqnarray}
\label{equ30}
\tilde{p}(\mu\,|\,x) & = & f_{S}(\mu\,|\,\mu \in [-\varepsilon, \varepsilon],x)\hspace{0.05em}
\tilde{p}(\mu \in [-\varepsilon, \varepsilon]\,|\,x) \nonumber \\ & & +
\hspace{0.2em}
f_{S}(\mu\,|\,\mu \notin [-\varepsilon, \varepsilon],x)\hspace{0.05em} \tilde{p}(\mu \notin
[-\varepsilon, \varepsilon]\,|\,x)
\end{eqnarray}
where the densities $f_{S}(\mu\,|\,\mu \in [-\varepsilon, \varepsilon],x)$ and
$f_{S}(\mu\,|\,\mu \notin [-\varepsilon, \varepsilon],x)$ are obtained by conditioning the density
function of $\mu$ given by the expression in equation~(\ref{equ29}) on the requirement that $\mu
\in [-\varepsilon, \varepsilon]$ and that $\mu \notin [-\varepsilon, \varepsilon]$, respectively.

However, we will now take the opportunity to deviate slightly from the methodology that was
outlined in the previous section by defining the marginal fiducial density of $\mu$ conditional on
$\mu$ lying in the interval $[-\varepsilon, \varepsilon]$ in the following way:
\begin{equation}
\label{equ31}
f_{*}(\mu\,|\,\mu \in [-\varepsilon, \varepsilon],x) = \mathtt{C}_7
(1 + \tau h(\mu)) f_{S}(\mu\,|\,\mu \in [-\varepsilon, \varepsilon],x)
\end{equation}
where $\tau \geq 0$ is a given constant, the function $h(\mu)$ is defined as the function
$h(\theta_j)$ was defined in Section~\ref{sec7}, the density function $f_{S}(\mu\,|\,\mu \in
[-\varepsilon, \varepsilon],x)$ is as has just been defined and $\mathtt{C}_7$ is a
normalising constant.
Here we are therefore, in effect, using the same kind of approach as put forward in
Section~\ref{sec7}, but applying this approach to the marginal fiducial density of the parameter
$\theta_j$, i.e.\ the parameter $\mu$ in the present case, rather than the fiducial density of
$\theta_j$ conditional on the parameters $\theta_{-j}$.

To clarify, by redefining the marginal fiducial density of $\mu$ in this way, we are also
redefining the post-data probability $\tilde{p}(\mu \in [-\varepsilon, \varepsilon]\,|\,x)$ that
appears in equation~(\ref{equ30}). This is a consequence of the joint fiducial density
$f(\theta\,|\,\theta^{A},x)$ that appears in equation~(\ref{equ24}), i.e.\ the joint density
$f(\mu, \sigma\,|\,\mu \in [-\varepsilon, \varepsilon],x)$ in the present case, being naturally
redefined as follows:
\pagebreak
\[
f_{*}(\mu, \sigma\,|\,\mu \in [-\varepsilon, \varepsilon],x) =
f_{*}(\mu\,|\,\mu \in [-\varepsilon, \varepsilon],x)f_{S}(\sigma\,|\,\mu,x)
\vspace{1ex}
\]
where the conditional fiducial density $f_{S}(\sigma\,|\,\mu,x)$ is as defined in
equation~(\ref{equ21}).

It will be assumed that the constant $\tau$ in equation~(\ref{equ31}) is determined in a similar
way to how we assumed the constant $\tau$ in equation~(\ref{equ16}) is determined when the
meth\-od\-ol\-ogy outlined in Section~\ref{sec7} was developed.
In particular, taking into account that the marginal post-data density $\tilde{p}(\mu\,|\,x)$ is
defined by equation~(\ref{equ30}) but with the fiducial density $f_{S}(\mu\,|\,\mu \in
[-\varepsilon, \varepsilon],x)$ in this equation now replaced by the fiducial density
$f_{*}(\mu\,|\,\mu \in [-\varepsilon, \varepsilon],x)$, we will assume that the constant $\tau$ is
chosen in a way that implies that the marginal density $\tilde{p}(\mu\,|\,x)$ will be continuous in
the vicinity of the points $\mu = -\varepsilon$ and $\mu = \varepsilon$.

To illustrate the use of the aforementioned calculations, Figure~8(a) shows plots of the marginal
post-data density $\tilde{p}(\mu\,|\,x)$ that has just been specified for various values of the
observed mean $\bar{x}$.
As was the case both earlier in this section and in Section~\ref{sec14}, it has been assumed that
the sample size $n$ equals 9 and the sample standard deviation $s$ equals 3.
Also, similar to the examples discussed in Sections~\ref{sec10} and~\ref{sec14}, it has been
assumed that $\varepsilon=0.2$ and the prior probability $p(\mu \in [-\varepsilon,\varepsilon])$ is
equal to 0.33.
Finally, as was the case in Figure~5, the short-dashed, solid and long-dashed curves in
Figure~8(a) correspond to the observed mean $\bar{x}$ being equal to 1.7, 2.1 and 2.5,
respectively.

\begin{figure}[t]
\begin{center}
\makebox[\textwidth]{\includegraphics[width=7.7in]{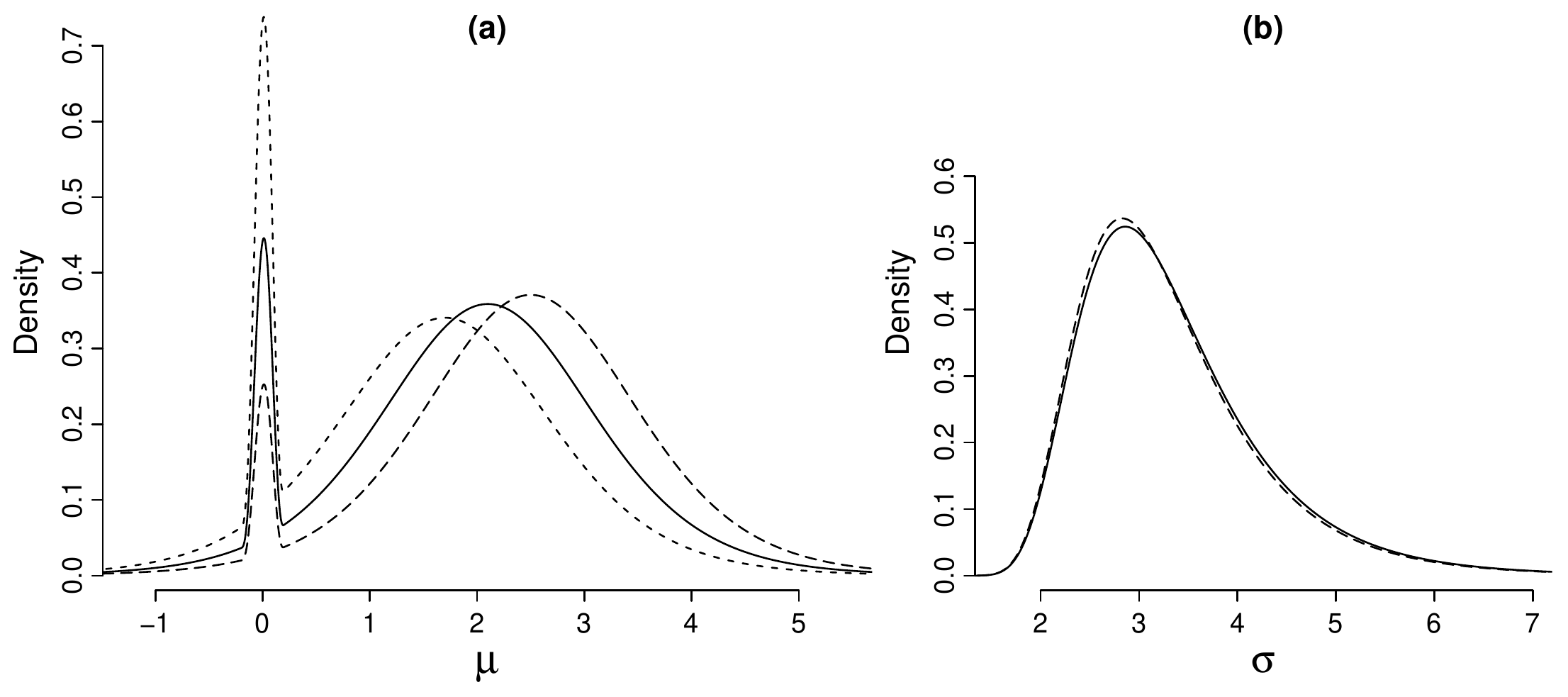}}
\caption{\small{Marginal post-data densities of the mean $\mu$ and the standard deviation $\sigma$
of a normal distribution in the case where the sample standard deviation $s$ equals 3 and the
sample size $n$ equals 9}}
\end{center}
\end{figure}

The solid curve in Figure~8(b) is a plot of the marginal density of $\sigma$ over the joint density
of $\mu$ and $\sigma$ that is natural to regard as being the joint post-data density of $\mu$ and
$\sigma$ in this example, i.e.\ the density function:
\[
\tilde{p}(\mu,\sigma\,|\,x) = \tilde{p}(\mu\,|\,x) f_{S}(\sigma\,|\,\mu,x)
\]
where the marginal post-data density $\tilde{p}(\mu\,|\,x)$ is defined by equation~(\ref{equ30})
but with the density $f_{S}(\mu\,|\,\mu \in [-\varepsilon, \varepsilon],x)$ in this equation
replaced by the density $f_{*}(\mu\,|\,\mu \in [-\varepsilon, \varepsilon],x)$ as specified by
equation~(\ref{equ31}), and where the conditional fiducial density $f_{S}(\sigma\,|\,\mu,x)$ is as
defined in equation~(\ref{equ21}).
This plot is based on the assumption that the sample mean $\bar{x}$ is equal to 2.1, however if
similar plots for the cases where $\bar{x}$ equals 1.7 and 2.5 had been added to Figure~8(b), then
although the marginal post-data densities of $\sigma$ for these three cases are different from each
other, they are so similar that this would have only really led to a fattening of the solid line
that represents the case where \pagebreak $\bar{x}=2.1$.
All other assumptions about quantities of interest are the same as were made in constructing the
plots in Figure~8(a).
To provide a contrast to the single solid curve that is plotted in Figure~8(b), a long-dashed curve
has been added to this figure that represents the marginal fiducial density $f_{S}(\sigma\,|\,x)$
defined by the expression in equation~(\ref{equ32}). This fiducial density of $\sigma$, which of
course does not depend on the value of the sample mean $\bar{x}$, is also represented by the solid
curve in Figure~6(b).

In the case where the sample mean $\bar{x}$ equals 2.1, it can be appreciated from comparing the
marginal post-data density of $\mu$ represented by the histogram in Figure~6(a) with the marginal
post-data density of $\mu$ represented by the solid curve in Figure~8(a) that if making
unconditional inferences about the mean $\mu$ is our main objective then, from a practical point of
view, there is not that much difference in the results obtained by applying the method outlined in
Section~\ref{sec15} from applying the method outlined in Section~\ref{sec12}.
In this respect, it is of interest to note that, for the case in question, the marginal post-data
probability that the special hypothesis about $\mu$ will be true, or in other words, that $\mu$
will lie in the interval $[-0.2,0.2]$, is equal to 0.105 and 0.092 in applying each of the two
methods put forward in Sections~\ref{sec12} and~\ref{sec15}, respectively.
The difference between these two post-data probabilities is arguably relatively small in the
context of the imprecision there is likely to be in many practical situations in determining the
most appropriate prior probability to assign to the hypothesis in question.
Furthermore, in an additional analysis, a similar relative difference in the marginal post-data
probability of $\mu$ lying in the interval $[-0.2,0.2]$ was observed when this probability was
derived separately using each of the two methods in question for a wide range of alternative values
for the sample mean $\bar{x}$, i.e.\ from zero up to very large values of $\bar{x}$.

\vspace{3ex}
\section{Sixth example: Inference about a relative risk}
\label{sec18}

To go, in a certain sense, completely around the circle of examples considered in the present
paper, let us now apply the method of inference that was put forward in Section~\ref{sec15} to the
problem of inference that was posed in Example~1.2 of the Introduction, i.e.\ that of making
inferences about a relative risk $\pi_t/\pi_c$ of a given adverse event of interest.

In doing this, it will be generally assumed that the proportions $\pi_t$ and $\pi_c$ are both
unknown. Also, for the same reason as given in the description of the example in question, we will
suppose that the scenario of Definition~2 would apply if the parameter $\nu$ is taken as being the
relative risk $\pi_t/\pi_c$ and the special hypothesis is taken as being the hypothesis that
$\pi_t/\pi_c$ lies in the interval $[1/(1+\varepsilon), 1 + \varepsilon]$, where $\varepsilon$ is
again a given small non-negative constant.
To be more precise, this will be assumed to be the case independent of whether $\pi_t$ and $\pi_c$
are both unknown, whether $\pi_c$ became known while $\pi_t$ remained unknown or whether $\pi_t$
became known while $\pi_c$ remained unknown. We should also make clear that the number of events in
the treatment and control groups, i.e.\ the observed counts $e_t$ and $e_c$ in terms of the
notation used in the Introduction, will be assumed to have binomial distributions.

Having clarified the general nature of the problem of interest, let us now furthermore suppose that
the fiducial densities $f(\theta\,|\,\theta^{A},x)$ and $f(\theta\,|\,\theta^{B},x)$ referred to in
Section~\ref{sec15}, i.e.\ the densities
$f(\pi_t, \pi_c\,|\, \pi_t/ \pi_c \in [1/(1+\varepsilon), 1+\varepsilon],x)$ and
$f(\pi_t, \pi_c\,|\, \pi_t/ \pi_c \notin [1/(1+\varepsilon), 1+\varepsilon],x)$ in the present
case, are defined by:\\[1ex]
$f(\pi_t, \pi_c\,|\, \pi_t/ \pi_c \in [1/(1+\varepsilon), 1+\varepsilon],x)$
\begin{flalign}
\label{equ35}
&& = \left\{
\begin{array}{ll}
\mathtt{C}_8 (1 + \tau h (\pi_t/ \pi_c))f_S(\pi_t\,|\,x)f_S(\pi_c\,|\,x)\ &
\mbox{if $\pi_t/ \pi_c \in [1/(1+\varepsilon), 1+\varepsilon]$}\\[1ex]
0 & \mbox{otherwise}
\end{array}
\right.
\end{flalign}
and by:\\[1ex]
$f(\pi_t, \pi_c\,|\, \pi_t/ \pi_c \notin [1/(1+\varepsilon), 1+\varepsilon],x)$
\begin{flalign}
\label{equ36}
&& = \left\{
\begin{array}{ll}
\mathtt{C}_9\hspace{0.05em} f_S(\pi_t\,|\,x)f_S(\pi_c\,|\,x)\ &
\mbox{if $\pi_t/ \pi_c \notin [1/(1+\varepsilon), 1+\varepsilon]$}\\[1ex]
0 & \mbox{otherwise}
\end{array}
\right.
\end{flalign}
\par \vspace{1.5ex} \noindent
where the data set $x$ is, of course, the set of counts $\{e_t,n_t,e_c,n_c\}$ referred to in
Example~1.2 of the Introduction, $\tau \geq 0$ is a given constant, $\mathtt{C}_8$ and
$\mathtt{C}_9$ are normalising constants, the function $h(\pi_t/ \pi_c)$ is defined as the function
$h(\theta_j)$ was defined in Section~\ref{sec7} and the fiducial densities $f_S(\pi_t\,|\,x)$ and
$f_S(\pi_c\,|\,x)$ are defined by the expressions:
\vspace{0.5ex}
\begin{equation}
\label{equ33}
f_S(\pi_t\,|\,x) = \int_{0}^{1} \omega_{t}(\pi_t \,|\,\gamma) d\gamma
\end{equation}
\begin{equation}
\label{equ34}
f_S(\pi_c\,|\,x) = \int_{0}^{1} \omega_{c}(\pi_c \,|\,\gamma) d\gamma
\end{equation}
\par \vspace{1ex} \noindent
in which
\vspace{1ex}
\[
\omega_{t}(\pi_t\,|\,\gamma) = \left\{
\begin{array}{ll}
\mathtt{C}_{10}(\gamma)\hspace{0.1em}\pi_t^{\hspace{0.05em}e_t}
(1-\pi_t)^{n_t-e_t}\, & \mbox{if $\pi_t \in \pi_t(\gamma)$}\\[1ex]
0 & \mbox{otherwise}
\end{array}
\right.
\]
and
\vspace{1ex}
\[
\omega_{c}(\pi_c\,|\,\gamma) = \left\{
\begin{array}{ll}
\mathtt{C}_{11}\hspace{-0.05em}(\gamma)\hspace{0.1em}\pi_c^{\hspace{0.05em}e_c}
(1-\pi_c)^{n_c-e_c}\, & \mbox{if $\pi_c \in \pi_c(\gamma)$}\\[1ex]
0 & \mbox{otherwise}
\end{array}
\right.
\]
\par \vspace{2.5ex} \noindent
where $\mathtt{C}_{10}(\gamma)$ and $\mathtt{C}_{11}\hspace{-0.05em}(\gamma)$ are normalising
constants that depend on the value of $\gamma$ and where $\pi_t(\gamma)$ and $\pi_c(\gamma)$ are
the sets of values of $\pi_t$ and $\pi_c$ that map onto the value $\gamma$ for the primary r.v.\
$\Gamma$ according to the equations $e_t = \varphi(\Gamma, \pi_t)$ and
$e_c = \varphi(\Gamma, \pi_c)$, respectively, under the assumption that the functions
$\varphi(\Gamma, \pi_t)$ and $\varphi(\Gamma, \pi_c)$ are defined in a \linebreak similar way to
how the function $\varphi(\Gamma, \pi)$ was defined in equation~(\ref{equ38}).

To make the reasoning behind the definitions that have just been presented easier to understand,
let us make the following comments:

\pagebreak
\noindent
1) The fiducial densities $f_S(\pi_t\,|\,x)$ and $f_S(\pi_c\,|\,x)$ defined in
equations~(\ref{equ33}) and~(\ref{equ34}) have the same form as the fiducial density
$f_S(\theta_j\,|\,\theta_{-j},x)$ in the example considered in Section~\ref{sec8}, i.e.\ the
fiducial density $f_S(\pi\,|\,x)$ according to the notation used in this earlier section.
Therefore, it would be quite reasonable to regard the fiducial densities $f_S(\pi_t\,|\,x)$ and
$f_S(\pi_c\,|\,x)$ as being suitable to use as post-data densities of $\pi_t$ and $\pi_c$ in a
general scenario where nothing or very little was known about the parameters $\pi_t$ and $\pi_c$
before the data set $x$ was observed.

\vspace{2ex}
\noindent
2) Since the fiducial density $f_S(\pi_t\,|\,x)$ does not depend on the value of $\pi_c$ and the
fiducial density $f_S(\pi_c\,|\,x)$ does not depend on the value of $\pi_t$, the joint fiducial
density of $\pi_t$ and $\pi_c$ that corresponds to the general scenario just mentioned is
naturally obtained by multiplying these two marginal fiducial densities of $\pi_t$ and $\pi_c$
together. Let us denote this joint density function as $f_S(\pi_t,\pi_c\,|\,x)$.

\vspace{2ex}
\noindent
3) As can be seen from equation~(\ref{equ36}), the fiducial density
$f(\pi_t, \pi_c\,|\, \pi_t/ \pi_c \notin [1/(1+\varepsilon), 1+\varepsilon],x)$ is obtained by
simply conditioning the joint fiducial density of $\pi_t$ and $\pi_c$ just referred to, i.e.\ the
density $f_S(\pi_t,\pi_c\,|\,x)$, on the requirement that $\pi_t/ \pi_c \notin [1/(1+\varepsilon),
1+\varepsilon]$, which is consistent with the assumptions that have been made with regard to the
scenario of interest.

\vspace{2ex}
\noindent
4) As can be seen from equation~(\ref{equ35}), the fiducial density
$f(\pi_t, \pi_c\,|\, \pi_t/ \pi_c \in [1/(1+\varepsilon), 1+\varepsilon],x)$ is effectively
obtained by applying the extension of Principle~2 outlined in Section~\ref{sec3} under the
assumption that the GPD function $\omega_G (\pi_t/ \pi_c)$ is as defined by equation~(\ref{equ16})
with the general parameter $\theta_j$ set equal to $\pi_t/ \pi_c$. However, by contrast to the
fiducial density $f(\theta_j\,|\,\theta_{-j},x)$ defined by equation~(\ref{equ7}), the fiducial
density $f(\pi_t, \pi_c\,|\, \pi_t/ \pi_c \in [1/(1+\varepsilon), 1+\varepsilon],x)$ is a joint
density function of two parameters rather than a univariate density function.

\vspace{2ex}
\noindent
5) It will be assumed that the constant $\tau$ in equation~(\ref{equ35}) is chosen in a way that
implies that the marginal density of $\pi_t/ \pi_c$ over the joint post-data density
$\tilde{p}(\pi_t, \pi_c\,|\,x)$ defined by equation~(\ref{equ37}), i.e.\ the marginal post-data
density $\tilde{p}(\pi_t / \pi_c\,|\,x)$, will be continuous in the vicinity of the points
$\pi_t/ \pi_c = 1/(1+\varepsilon)$ and $\pi_t/ \pi_c = 1+\varepsilon$.
Therefore, we are assuming that the constant $\tau$ will be chosen in a similar way to how we
assumed the constant $\tau$ in equation~(\ref{equ16}) would be determined whenever the methodology
outlined in Section~\ref{sec7} is applied.

\vspace{2ex}
Given the definitions of the fiducial densities $f(\pi_t, \pi_c\,\hspace{0.05em}|\,\hspace{0.05em}
\pi_t/ \pi_c\hspace{-0.05em} \in\hspace{-0.05em} [1/(1+\varepsilon), 1 + \varepsilon] ,x)$ and
$f(\pi_t, \pi_c\,\hspace{0.05em}|\,\hspace{0.05em} \pi_t/ \pi_c\hspace{-0.05em}
\notin\hspace{-0.05em} [1/(1+\varepsilon), 1 + \varepsilon] ,x)$ in equations~(\ref{equ35})
and~(\ref{equ36}), and bearing in mind that, similar to the example outlined in Section~\ref{sec8},
the sample counts of interest, i.e.\ $e_t$ and $e_c$, have binomial distributions, we can use
equation~(\ref{equ27}) to determine the post-data probability
$\tilde{p}(\theta = \theta^{A}\,|\,x)$ in the present case, i.e.\ the post-data probability
$\tilde{p}(\pi_t/ \pi_c \in [1/(1+\varepsilon), 1+\varepsilon]\,|\,x)$, for any given value that is
assigned to the prior probability that $\pi_t/ \pi_c$ lies in the interval $[1/(1+\varepsilon),
1+\varepsilon]$, i.e.\ the prior probability $p_0(\theta^A)$ in terms of the notation used in the
equation in question.
Having performed this calculation to determine the post-data probability being referred to, we can
then use equation~(\ref{equ37}) to determine the joint post-data density of the proportions $\pi_t$
and $\pi_c$ over all possible values of these two proportions, i.e.\ the post-data density
$\tilde{p}(\pi_t,\pi_c\,|\,x)$.

To illustrate the use of the calculations just discussed, Figure~9 shows some results from
generating values from the joint post-data density $\tilde{p}(\pi_t,\pi_c\,|\,x)$ that has just
been defined for the case where the observed counts are as follows: $e_t=6$, $n_t=20$, $e_c=18$ and
$n_c=30$. In particular, the histograms in Figures~9(a) to~9(c) represent, respectively, the
marginal density of $\pi_t$, the marginal density of $\pi_c$ and the marginal density of
$\pi_t/ \pi_c$ over the joint post-data density of $\pi_t$ and $\pi_c$ in question.
The general assumptions that were made to perform the calculations that underlie these figures were
that $\varepsilon=0.045$, the prior probability of $\pi_t/ \pi_c$ lying in the interval
$[1/(1+\varepsilon), 1+\varepsilon]$ is equal to 0.4 and the density function $h(\pi_t/ \pi_c)$
that appears in equation~(\ref{equ35}) is defined such that:
\[
\log(\pi_t/ \pi_c) \sim \mbox{Beta}\hspace{0.1em} (4,4,-\log(1+\varepsilon),\log(1+\varepsilon))
\]
where the notation used here is the same as used in equation~(\ref{equ17}), which means therefore
that the density $h(\pi_t/ \pi_c)$ is symmetrical on the logarithmic scale.
The numerical output on which the histograms in the figures in question are based was generated by
the method of importance sampling, more specifically by appropriately weighting an independent
sequence of four million random values of $\pi_t$ and $\pi_c$ drawn from the joint fiducial density
$f_S(\pi_t,\pi_c\,|\,x)$ that was referred to in the above discussion.

\begin{figure}[!t]
\begin{center}
\makebox[\textwidth]{\includegraphics[width=7.7in]{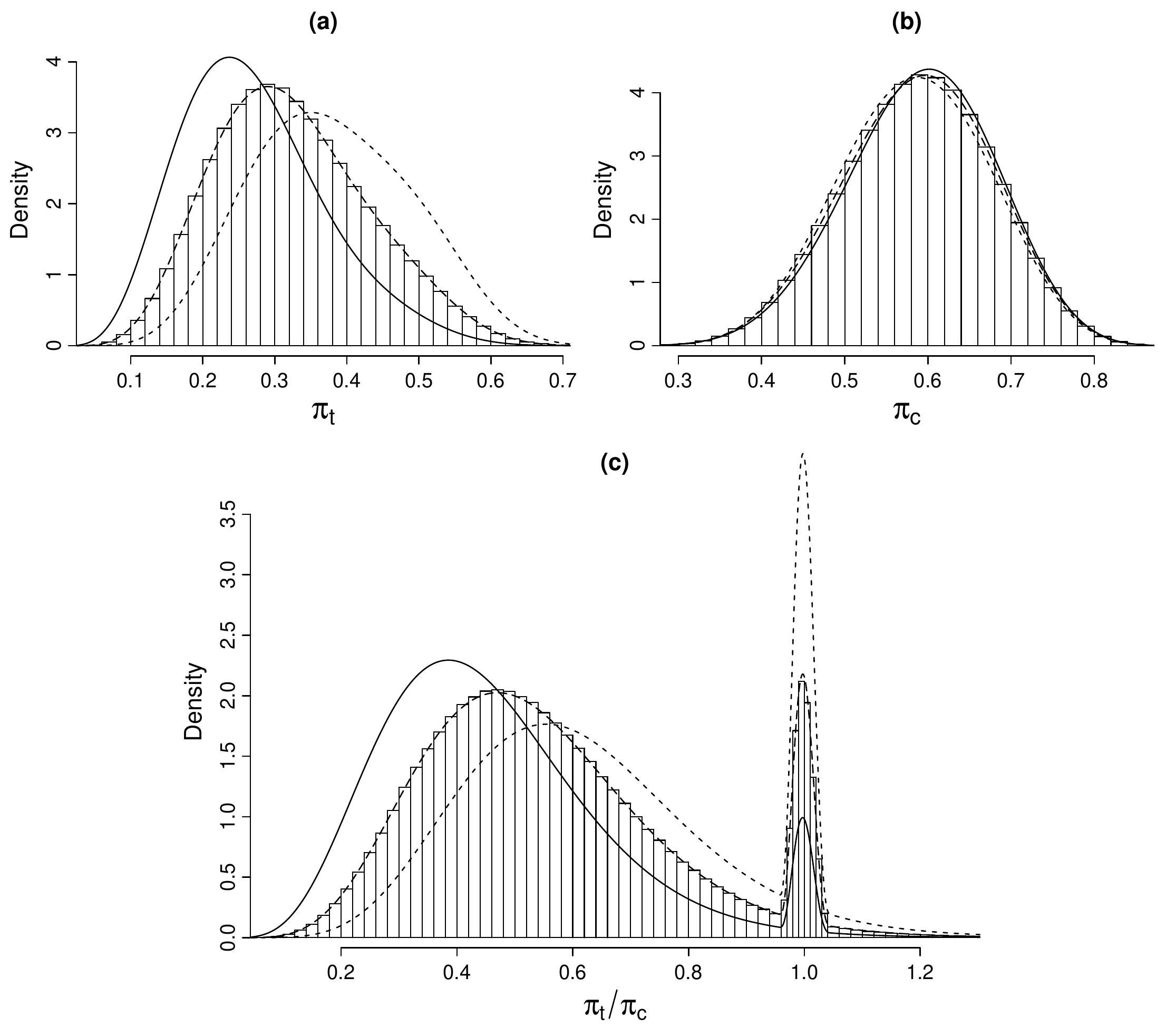}}
\par \vspace{1ex}
\caption{\small{Marginal post-data densities of the binomial proportions $\pi_t$ and $\pi_c$ and
of the relative risk $\pi_t / \pi_c$ for three different values of the observed count $e_t$,
namely 5, 6 and 7, in the case where $n_t=20$, $e_c=18$, $n_c=30$, $\varepsilon=0.045$ and the
prior probability that $\pi \in [0.957,1.045]$ is equal to 0.4}}
\end{center}
\end{figure}

\pagebreak
The curves overlaid on the histograms in Figures~9(a) to~9(c) are plots of approximations to the
marginal densities of $\pi_t$, $\pi_c$ and $\pi_t/ \pi_c$ over the joint post-data density
$\tilde{p}(\pi_t,\pi_c\,|\,x)$ that were obtained by substituting the fiducial densities
$f_S(\pi_t\,|\,x)$ and $f_S(\pi_c\,|\,x)$ that appear in equations~(\ref{equ35})
and~(\ref{equ36}) by the posterior densities of the parameters $\pi_t$ and $\pi_c$ that are based
on the Jeffreys prior density for these parameters, i.e.\ based on choosing the prior density for
each of the proportions $\pi_t$ and $\pi_c$ to be the beta density function of the proportion
concerned that has both its shape parameters equal to 0.5. A simulation study that was carried out
showed that this method of approximation was satisfactory for the range of values of $e_t$, $n_t$,
$e_c$ and $n_c$ being used in this example.
More specifically, the solid curves in Figures~9(a) to~9(c) represent, respectively, the marginal
post-data densities of $\pi_t$, $\pi_c$ and $\pi_t/ \pi_c$ just referred to in the case where the
observed counts are $e_t=5$, $n_t=20$, $e_c=18$ and $n_c=30$, while the short-dashed curves in
these figures represent the marginal post-data densities of $\pi_t$, $\pi_c$ and $\pi_t/ \pi_c$
just referred to in the case where $e_t=7$, $n_t=20$, $e_c=18$ and $n_c=30$.
Finally, the long-dashed curves in the figures in question represent the same marginal post-data
densities of $\pi_t$, $\pi_c$ and $\pi_t/ \pi_c$ that are represented by the histograms in these
figures, i.e.\ these curves correspond to the case where the observed counts are $e_t=6$, $n_t=20$,
$e_c=18$ and $n_c=30$. The fact that the histograms in Figures~9(a) to~9(c) are closely
approximated by the long-dashed curves in question is, of course, an indication of the quality of
the approximation that was used to construct these long-dashed curves.

It can be seen from Figure~9(c) that, for all three values of $e_t$ to which the curves in this
figure correspond, i.e.\ the values 5, 6 and 7, a substantial amount of the probability mass of the
marginal density of $\pi_t/ \pi_c$ over the joint post-data density $\tilde{p}(\pi_t,\pi_c\,|\,x)$
can be found in the interval $[1/(1+\varepsilon), 1+\varepsilon]$, or more specifically in the
interval $[0.957,1.045]$, but also it can be seen that the post-data probability that
$\pi_t/ \pi_c$ lies in this interval decreases sharply as the sample proportion $e_t/n_t$ moves
away from the sample proportion $e_c/n_c$.
To give some more detail here, under the assumption that $n_t=20$, $e_c=18$ and $n_c=30$, the
post-data probability being referred to, i.e.\ $\tilde{p}(\pi_t/ \pi_c \in [0.957,1.045]\,|\,x)$,
is approximately equal to 0.0424, 0.0927 and 0.168 when the event count $e_t$ is equal to 5, 6
and 7, respectively, which are probabilities that, as would naturally be expected, are smaller than
the prior probability of 0.4 that was assigned to the hypothesis that $\pi_t/ \pi_c$ would lie in
the interval in question.
Also, it should be pointed out that, although the marginal post-data density of $\pi_t$ in
Figure~9(a) that corresponds to the case where $e_t=7$, i.e.\ the density function represented by
the short-dashed curve in this figure, would appear to have a surprisingly bulbous form, this is
simply due to this density function of $\pi_t$ being heavily pulled towards the sample proportion
$e_c/n_c$, i.e.\ the value 0.6, relative to the form that this density function would have if the
prior probability of $\pi_t/ \pi_c$ lying in the interval $[0.957,1.045]$ had been chosen to be a
much smaller value.

\vspace{3ex}
\section{Discussion}
\label{sec17}

\vspace{0.5ex}
\noindent
\textbf{Utility and convenience of the proposed approach}

\vspace{1.5ex}
\noindent
The methodology put forward in the present paper allows marginal or joint post-data probability
densities to be constructed for the parameters of given models when there was a substantial degree
of pre-data or prior belief that one of the model parameters $\theta_j$ would equal, or lie close
to, a specific value, or in other words, that $\theta_j$ would lie in a given narrow interval
$[\theta_{j0}, \theta_{j1}]$, but apart from this very little or nothing was known about the
parameters concerned. To achieve this, usually the only input of major note that is required from
the user is the assignment of a prior probability to the hypothesis that the parameter $\theta_j$
lies in the interval $[\theta_{j0}, \theta_{j1}]$, which is a probability that may be either
conditional or unconditional on the values of the rest of the parameters $\theta_{-j}$. Also, we
should mention that, as discussed in Section~\ref{sec12}, the approach allows inferences to be made
about the parameters of a given model in the case where, for each of the parameters $\theta_j$ in a
given set of parameters $\theta^{(1)}$, there was a substantial degree of pre-data belief that the
parameter concerned would lie in a given narrow interval $[\theta_{j0}, \theta_{j1}]$.

\vspace{4ex}
\noindent
\textbf{Comparison with a method proposed by Aitken}

\vspace{1.5ex}
\noindent
For any given prior probability that is assigned to the hypothesis that the parameter $\theta_j$
lies in the interval $[\theta_{j0}, \theta_{j1}]$, the approach outlined in the present paper will
lead, in some examples, such as the example considered in Section~\ref{sec6}, to the same post-data
probability being assigned to this hypothesis as would be the case when applying the method that
was proposed in Aitken~(1991) which is based on what are called `posterior Bayes factors'.
Given that at the time this paper was published, this method faced heavy criticism, as can be seen
from the comments in the published discussion that immediately follows the paper in question, it is
of interest to know how the methodology put forward in the present paper stands up to, or avoids,
these criticisms.

We may first observe that the method outlined in Aitken~(1991) is based on the first type of
Bayesian analogy that was considered in Section~\ref{sec4}, i.e.\ the analogy that corresponds to a
full Bayesian analysis, and therefore inherits all the difficulties associated with trying to
express our pre-data knowledge about the parameter of interest $\theta_j$ in the form of a prior
density function over $\theta_j$ when we are in the scenario of Definition~2. As has already been
discussed, it is difficult to `sweep this issue under the carpet' given that the posterior
probability of $\theta_j$ lying in the interval $[\theta_{j0}, \theta_{j1}]$ will generally be
highly sensitive to the choice made for the prior density of $\theta_j$ when it has already been
decided that a given prior probability mass must lie in the interval $[\theta_{j0}, \theta_{j1}]$.

In trying to combat this issue, a second poor decision was arguably made in constructing the method
proposed in Aitken~(1991), which was, in effect, to stipulate that, conditional either on
$\theta_j$ lying in the interval $[\theta_{j0}, \theta_{j1}]$ or on $\theta_j$ lying outside of
this interval, inferences should be made about the parameter $\theta_j$ by placing an `objective'
prior density over $\theta_j$, and in particular, the paper in question effectively endorses
specifying these conditional prior densities of $\theta_j$ in a way that depends on the sampling
model, e.g.\ it endorses the use of the Jeffreys prior. Therefore, the method betrays the Bayesian
analogy on which it is based, and therefore, as discussed for example in Bowater~(2020), is
arguably neither a Bayesian method nor a method that is easy to justify directly in some other way.

Finally, even if we accept the use of the type of model-based prior densities that were just
mentioned, the method advocated in Aitken~(1991) does not even make a correct use of Bayes' theorem
as a purely abstract mathematical formula. More specifically, in forming a supposed unconditional
`posterior' density of $\theta_j$, the method does not combine the likelihood function with the
unconditional prior density of $\theta_j$ that corresponds to the conditional prior densities and
marginal prior probabilities on which the method is based in the way that is correct according to
Bayes' theorem. This defect is repeatedly highlighted in the published discussion that immediately
follows Aitken~(1991).

It is also worth pointing out that as an attempt to correct for a specific element of incoherency
that was identified in the method outlined in Aitken~(1991), an alternative method was proposed in
O'Hagan~(1995) which attempts to achieve the same objective that was pursued in Aitken~(1991) by
using what are called `fractional Bayes factors'. Although, as can be seen from the comments in
the published discussion that immediately follows O'Hagan~(1995), the evaluation of this
alternative method was, at the time, markedly more favourable than the evaluation of the method
that was earlier proposed in Aitken~(1991), it is unclear what the justifiable reason for this
actually is, since it is in fact open to exactly the same criticisms that were just made about this
earlier method.

By contrast, the method put forward in the present paper avoids all of the criticisms in question
by being based on making the second Bayesian analogy that was outlined in Section~\ref{sec4} rather
than the first one, and by ensuring that all the methodological elements that are introduced into
this approach are consistent with this alternative analogy.
No excuses will be made here for accepting the point of view that the most, if not the only,
appropriate way of using Bayes' theorem to tackle any given problem of statistical inference is by
making some kind of analogy between the uncertainty that surrounds the validity of hypotheses that
are relevant to this problem and the uncertainty that surrounds the outcomes of the type of
physical experiment to which Bayes' theorem is most naturally applied.

\vspace{4ex}
\noindent
\textbf{Extending the approach to deal with general model comparison}

\vspace{1.5ex}
\noindent
To conclude this paper, let us imagine that there are a set of $r$ possible models $\mathcal{M} =
\{M_i:i=1,2,\ldots,r\}$ that could have generated a data set of interest $x$ and that each of these
models $M_i$ is fully specified by a set of unknown parameters
$\theta^{[\hspace{0.05em}i\hspace{0.05em}]}$. Also, we will suppose that conditional on any given
one of these models $M_i$ being the true model, very little or nothing would have been known about
the parameter values in the set $\theta^{[\hspace{0.05em}i\hspace{0.05em}]}$ before the data were
observed.
In these circumstances, the approach outlined in the present paper could be extended such that, for
any given prior probabilities that are assigned to the models in the set $\mathcal{M}$, it would be
possible to determine post-data probabilities of each of these models being true, and as a result,
construct a post-data probability distribution over the space of all the models concerned.
It is clear that some of the examples discussed in the present paper would naturally form special
cases of such a general framework and that the use of this framework may well allow us to overcome
difficulties in the general use of standard Bayes factors that were identified, for example, in
Aitken~(1991), O'Hagan~(1995) and Berger and Pericchi~(1996). Nevertheless, we will choose to end
the present paper at this point with the hope that attentive readers may be able to flesh out the
details of such a general framework by themselves.

\vspace{5.5ex}
\noindent
\textbf{References}

\begin{description}

\setlength{\itemsep}{1ex}

\vspace{0.5ex}
\item[] Aitken, M. (1991).\ Posterior Bayes factors (with discussion).\ \emph{Journal of the Royal
Statistical Society, Series B}, \textbf{53}, 111--142.

\item[] Berger, J. O. and Pericchi, L. R. (1996).\ The intrinsic Bayes factor for model selection
and prediction.\ \emph{Journal of the American Statistical Association}, \textbf{91}, 109--122.

\item[] Berger, J. O. and Sellke, T. (1987).\ Testing a point null hypothesis: the
irreconcilability of P values and evidence.\ \emph{Journal of the American Statistical
Association}, \textbf{82},\linebreak 112--122.

\item[] Bowater, R. J. (2017).\ A defence of subjective fiducial inference.\ \emph{AStA Advances in
Statistical Analysis}, \textbf{101}, 177--197.

\item[] Bowater, R. J. (2018).\ Multivariate subjective fiducial inference.\ \emph{arXiv.org
(Cornell University), Statistics}, arXiv:1804.09804.

\item[] Bowater, R. J. and Guzm\'{a}n-Pantoja, L. E. (2019a).\ Bayesian, classical and hybrid
methods of inference when one parameter value is special.\ \emph{Journal of Applied Statistics},
\textbf{46}, 1417--1437.

\item[] Bowater, R. J. (2019b).\ Organic fiducial inference.\ \emph{arXiv.org (Cornell University),
Sta\-tis\-tics}, arXiv:1901.08589.

\item[] Bowater, R. J. (2019c).\ Sharp hypotheses and bispatial inference.\ \emph{arXiv.org
(Cornell University), Statistics}, arXiv:1911.09049.

\item[] Bowater, R. J. (2020).\ Integrated organic inference (IOI):\ a reconciliation of
statistical paradigms.\ \emph{arXiv.org (Cornell University), Statistics}, arXiv:2002.07966.

\item[] Bowater, R. J. (2021a).\ A very short guide to IOI:\ a general framework for statistical
inference summarised.\ \emph{arXiv.org (Cornell University), Statistics}, arXiv:2104.11766.

\item[] Bowater, R. J. (2021b).\ A revision to the theory of organic fiducial inference.\
\emph{arXiv.org (Cornell University), Statistics}, arXiv:2111.09279.

\item[] Bowater, R. J. (2022).\ Physical, subjective and analogical probability.\ \emph{arXiv.org
(Cornell University), Statistics}, arXiv:2204.10159.

\item[] Brooks, S. P. and Roberts, G. O. (1998).\ Convergence assessment techniques for Markov
chain Monte Carlo.\ \emph{Statistics and Computing}, \textbf{8}, 319--335.

\item[] Gelfand, A. E. and Smith, A. F. M. (1990).\ Sampling-based approaches to calculating
marginal densities.\ \emph{Journal of the American Statistical Association}, \textbf{85}, 398--409.

\item[] Gelman, A. and Rubin, D. B. (1992).\ Inference from iterative simulation using multiple
sequences.\ \emph{Statistical Science}, \textbf{7}, 457--472.

\item[] Geman, S. and Geman, D. (1984).\ Stochastic relaxation, Gibbs distributions and the
Bayesian restoration of images.\ \emph{IEEE Transactions on Pattern Analysis and Machine
Intelligence}, \textbf{6}, 721--741.

\item[] O'Hagan, A. (1995).\ Fractional Bayes factors for model comparison (with discussion).\
\emph{Journal of the Royal Statistical Society, Series B}, \textbf{57}, 99--138.

\end{description}

\end{document}